\documentclass[a4paper,11pt,headings=big,DIV=12]{scrartcl}
\pdfoutput=1
\usepackage{amsmath,amssymb}
\usepackage{color,xcolor}
\definecolor{blue}{rgb}{0,0,0.5}
\usepackage[colorlinks,linkcolor=blue,urlcolor=blue,citecolor=blue]{hyperref}
\usepackage{graphicx}
\addtokomafont{disposition}{\rmfamily\boldmath}
\usepackage{cite}
\usepackage{xspace}
\usepackage{relsize}
\usepackage{bm}
\usepackage{multirow}
\usepackage[utf8]{inputenc}

\allowdisplaybreaks

\begin{document}

\begin{center}
{\LARGE\bfseries \boldmath
\vspace*{1.5cm}
Constraining new physics in $b\to c\ell\nu$ transitions
}\\[0.8 cm]
{\large\textsc{
Martin Jung \& David M. Straub
}\\[0.5 cm]
\small
Excellence Cluster Universe, Boltzmannstr.~2, 85748~Garching, Germany
}
\\[0.5 cm]
\small
E-Mail: \texttt{
\href{mailto:martin.jung@tum.de}{martin.jung@tum.de}, \href{mailto:david.straub@tum.de}{david.straub@tum.de}}
\\[0.2 cm]
\end{center}

\bigskip

\begin{abstract}\noindent
$B$ decays proceeding via $b\to c\ell\nu$ transitions with $\ell=e$ or $\mu$ are tree-level processes in the Standard Model. They are
used to measure the CKM element $V_{cb}$, as such forming an important ingredient in the determination of \emph{e.g.}\ the unitarity
triangle;  hence the question to which extent they can be affected by new physics contributions is important, specifically given the
long-standing tension between $V_{cb}$ determinations from inclusive and exclusive decays and the significant hints for lepton flavour
universality violation in $b\to c\tau\nu$ and $b\to s\ell\ell$ decays. We perform a comprehensive model-independent analysis of new
physics in $b\to c\ell\nu$, considering all combinations of scalar, vector and tensor interactions occuring in single-mediator
scenarios. We include for the first time differential distributions of $B\to D^*\ell\nu$ angular observables for this
purpose.
We show that these are valuable in constraining non-standard interactions. Specifically, the zero-recoil endpoint of the $B\to D\ell\nu$ spectrum is extremely sensitive to scalar currents, while
the maximum-recoil endpoint of the $B\to D^*\ell\nu$ spectrum with transversely polarized $D^*$ is extremely sensitive to tensor
currents.
We also quantify the room for $e$-$\mu$ universality violation in $b\to c\ell\nu$ transitions, predicted by some models suggested to
solve the $b\to c\tau\nu$ anomalies, from a global fit to $B\to D\ell\nu$ and $B\to D^*\ell\nu$ for the first time. Specific new
physics models, corresponding to all possible tree-level mediators, are also discussed. As a side effect, we present $V_{cb}$
determinations from exclusive $B$ decays, both with frequentist and Bayesian statistics, leading to compatible results.
The entire numerical analysis is based on open source code, allowing it to be easily adapted once new data or new form factors become
available.
\end{abstract}

\newpage
\section{Introduction}

In the context of the Standard Model (SM), the element $V_{cb}$ of the Cabibbo-Kobayashi-Maskawa (CKM) matrix
can be determined in various ways:
\begin{itemize}
 \item from a global fit including \emph{e.g.}\ meson-antimeson mixing observables \cite{Charles:2015gya,Bona:2017cxr},
 \item from inclusive measurements of $B\to X_ce\nu$ \cite{Alberti:2014yda}, where $X_c$ is any charmed
 hadronic final state,
 \item from exclusive $b\to c\ell\nu$ transitions, specifically $B\to D\ell\nu$ and $B\to D^*\ell\nu$ ($\ell=e,\mu$) decays (see e.g.\
 \cite{Ricciardi:2017lne} for a recent review).
\end{itemize}
When considering the SM as the low-energy limit of a more fundamental theory of ``new physics'' (NP),
these determinations are  not applicable model-independently in general. Specifically, flavour-changing neutral current
processes like meson-antimeson mixing could easily be affected by NP, invalidating the global fit. In this case, a potential
disagreement between the global fit and the tree-level determinations from semi-leptonic $B$ decays
can signal the presence of NP.
However, even the tree-level processes could in principle be significantly affected by NP.
This possibility has been considered in the past in particular in view of the
long-standing tensions between $V_{cb}$ from different decay channels.
Clearly, these tensions could be due to statistical fluctuations
or underestimated theoretical uncertainties.
While the inclusive decay can be computed to high precision in an expansion in $\alpha_s$ and $1/m_{c,b}$, see \emph{e.g.}
\cite{Alberti:2014yda} for a recent overview, the exclusive decays require the knowledge of hadronic form factors. Lattice QCD (LQCD)
calculations of $B\to D$ form factors are now available also at non-zero recoil \cite{Na:2015kha,Lattice:2015rga} for the relevant SM
operators, but for $B\to D^*$ a full lattice calculation at non-zero recoil is still lacking \cite{Bailey:2014tva,Harrison:2017fmw}.
In both cases, the dependence on the chosen form factor parametrization has received considerable interest recently
\cite{Bigi:2016mdz,Bigi:2017njr,Bigi:2017jbd,Aoki:2016frl,Lattice:2015rga,Bernlochner:2017jka,Grinstein:2017nlq,Bernlochner:2017xyx,Jaiswal:2017rve},
indicating that at least part of the tension between inclusive and exclusive decays might stem from an underestimation of the
systematic or theoretical uncertainties.
Nevertheless, entertaining the possibility of NP as the origin of the tensions between SM and data is certainly worthwhile, since these
analyses are suggestive, but do not provide proof that the form factor parametrization is indeed the reason for the observed tension.

Additional interest in NP modifying $b\to c\ell\nu$ with light leptons
was generated by the signficant deviations from SM expectations in decays
based on the $b\to c\tau\nu$ transition, including
in $B\to D\tau\nu$, $B\to D^*\tau\nu$, and, most recently,
$B_c\to J/\psi\tau\nu$
\cite{Lees:2013uzd,Huschle:2015rga,Aaij:2015yra,Sato:2016svk,Hirose:2016wfn,Aaij:2017uff,Aaij:2017tyk}.
Many NP models have been proposed to explain these tensions.
Depending on the flavour structure of the model, $b\to c\ell\nu$ with light
leptons can be affected as well. This is true in particular for models
explaining simultaneously the apparent deviations from lepton flavour universality (LFU)
in $B\to K\ell\ell$ and $B\to K^*\ell\ell$, with $\ell=e$ or $\mu$
(see \emph{e.g.}\ \cite{Bauer:2015knc,Becirevic:2016oho,Cai:2017wry,Buttazzo:2017ixm}).
Hence an important question is to what extent LFU is tested in $b\to c\ell\nu$,
independent of any tension between $b\to c\ell\nu$ data and SM predictions.

Recent analyses of NP in $b\to c\ell\nu$  include
\cite{Dassinger:2007pj,Dassinger:2008as,Crivellin:2009sd,Feger:2010qc,Faller:2011nj,Crivellin:2014zpa,Colangelo:2016ymy}.
Most of them have focused on individual operators or specific subsets and only used experimental information from the measurements of
total branching ratios in exclusive decays.
Recently however, differential distributions of exclusive measurements,
including angular observables in $B\to D^*\ell\nu$, have been released by
the BaBar and Belle collaborations \cite{Aubert:2009ac,Dungel:2010uk,Glattauer:2015teq,Abdesselam:2017kjf}.
As will be shown in section~\ref{sec:np}, these data contain valuable information
that allows to independently
constrain different types of non-standard interactions in $b\to c\ell\nu$.
The main aim of this paper is to perform a comprehensive model-independent
analysis of all possible types of NP effects in $b\to c\ell\nu$, making use
of the wealth of experimental data.

This  paper is organized as follows:
In section~\ref{sec:heff}, the effective Hamiltonian for
$b\to c\ell\nu$ is defined
and relations among the operators implied by SM gauge invariance are discussed.
In section~\ref{sec:ff}, we discuss our treatment of $B\to D$ and $B\to D^*$
form factors that are  crucial ingredients in the analysis of exclusive $b\to c\ell\nu$ decays.
In section~\ref{sec:Vcb}, we perform fits to $V_{cb}$ from $B\to D\ell\nu$
and $B\to D^*\ell\nu$.
Reproducing the values in the literature, this step is useful as a cross-check
of our numerics. We also perform the analysis with frequentist and Bayesian
statistics, explicitly demonstrating their agreement.
In section~\ref{sec:np}, we perform the NP analyses, starting with a discussion
of NP models with tree-level mediators and their characteristic patterns
of Wilson coefficients, and subsequently discussing each of the relevant operator combinations.
Section~\ref{sec:conc} contains our conclusions.

An important feature of our analysis is that it is entirely based on
open-source code. We have implemented all observables of interest
as well as our predictions for $B\to D$ and $B\to D^*$ form factors
in the \texttt{flavio} flavour physics package \cite{Straub:2018kue}.
This has three benefits: First, it makes our analysis transparent and reproducible.
Second, it allows anyone to update the best-fit values of $V_{cb}$ or the
allowed ranges for the Wilson coefficients when new experimental data
or new lattice form factor computations become available.
Third, it easily allows to study the viability of more involved NP models with
multiple Wilson coefficients, that cannot be easily visualized in two-dimensional
plots.
Additionally, to corroborate the reliability of our results,
we have obtained all numerical results with a completely independent
Mathematica code.

\section{Effective Hamiltonian}\label{sec:heff}

The effective Hamiltonian for $b\to c\ell\nu$ transitions can be written as\footnote{We do not consider scenarios with light
right-handed neutrinos. This general form for NP in $b\to c\ell \nu$ transitions has been first considered in
Ref.~\cite{Goldberger:1999yh}.}
\begin{equation}
\mathcal H_\text{eff}^{b\to c\ell\nu_{\ell'}} = \frac{4 G_F}{\sqrt{2}} V_{cb} \left(O_{V_L}^{\ell\ell'} \delta_{\ell\ell'} + \sum_i C_i^{\ell\ell'} O_i^{\ell\ell'} + \text{h.c.}\right),
\end{equation}
where the sum runs over the following operators:
\begin{align}
O_{V_L}^{\ell\ell'} &= (\bar c_L \gamma^\mu b_L)(\bar \ell_L \gamma_\mu \nu_{\ell' L}) \,, &
O_{S_R}^{\ell\ell'} &= (\bar c_L b_R)(\bar \ell_R \nu_{\ell' L}) \,, &
O_T^{\ell\ell'} &= (\bar c_R \sigma^{\mu\nu} b_L)(\bar \ell_R \sigma_{\mu\nu}\nu_{\ell' L}) \,,
\nonumber\\
O_{V_R}^{\ell\ell'} &= (\bar c_R \gamma^\mu b_R)(\bar \ell_L \gamma_\mu \nu_{\ell' L}) \,, &
O_{S_L}^{\ell\ell'} &= (\bar c_R b_L)(\bar \ell_R \nu_{\ell' L}) \,,
\label{eq:ops}
\end{align}
with in general charged-lepton- \emph{and} neutrino-flavour-dependent coefficients.
Since we are focusing on decays with light leptons in the final state, we only consider $\ell=e$ or $\mu$,
but allow for $\ell'=e,\mu,\tau$, which cannot be distinguished experimentally.
We have defined the coefficients $C_i^{\ell\ell'}$ in \eqref{eq:ops} such that they vanish in the SM.
A lepton-flavour universal and diagonal NP effect in $C_{V_L}^{\ell\ell'}$ can always be absorbed by a shift in $V_{cb}$, since $V_{cb}$ is a free
parameter in the SM and presently not meaningfully constrained by CKM unitarity.
In the following, we will use the shorthands
\begin{equation}
C_i^\ell\equiv C_i^{\ell\ell} \,,\qquad
\tilde V_{cb}^\ell = V_{cb} (1+C_{V_L}^\ell) \,,\qquad \tilde C_i^\ell = C_i^{\ell}/(1+C_{V_L}^\ell)\,,
\end{equation}
where appropriate.

The operators in the effective Hamiltonian arise from more funadamental interactions at or above the electroweak scale. The
available high-energy data from LHC indicate the existence of another energy gap between the electroweak scale and that of NP.
In such a scenario interactions beyond the SM can be cast into another effective theory, with operators symmetric under the full SM
gauge group. For linearly realized electroweak symmetry breaking this effective theory is called \emph{Standard Model effectivie
field theory} (SMEFT), the operators of which can be ordered in terms of their mass dimension, with those at dimension six giving the
dominant contributions here \cite{Buchmuller:1985jz,Grzadkowski:2010es}.
The tree-level matching of SMEFT operators onto
the effective Hamiltonian \eqref{eq:ops} reads~\cite{Cirigliano:2012ab,Alonso:2015sja,Cata:2015lta,Aebischer:2015fzz}
\begin{align}
C_{V_L}^{\ell\ell'} &= -v^2 \,\frac{V_{ci}}{V_{cb}}\, C_{lq}^{(3)\ell\ell'i3} +v^2 \,\frac{V_{ci}}{V_{cb}}\,C_{\phi q}^{(3)i3}\,\delta_{\ell\ell'}
\,,&
C_{V_R}^{\ell\ell'} &= \frac{v^2}{2} \,C_{\phi ud}^{23} \,\delta_{\ell\ell'}
\,,\label{eq:smeftV}\\
C_{S_R}^{\ell\ell'} &= -\frac{v^2}{2} \,\frac{V_{ci}}{V_{cb}}\, C_{ledq}^{\ell\ell'3i}
\,,&
C_{S_L}^{\ell\ell'} &= -\frac{v^2}{2} \,\frac{V_{ci}}{V_{cb}}\, C_{lequ}^{(1)\ell\ell'3i}
\,,\label{eq:smeftS}\\
C_T^{\ell\ell'} &= -\frac{v^2}{2} \,\frac{V_{ci}}{V_{cb}} \,C_{lequ}^{(3)\ell\ell'3i}\,,
\label{eq:smeftT}
\end{align}
where the definition of the SMEFT operators can be found in \cite{Grzadkowski:2010es}\footnote{%
We use a normalization $\mathcal L_\text{SMEFT}=\sum_iC_i O_i$, i.e. dimension-6 operators have dimensions of inverse mass squared.}
and we are employing a weak basis for SMEFT where down-type and charged-lepton mass matrices are diagonal.
An important prediction of this framework is that
the Wilson coefficient $C_{V_R}^{\ell\ell'}$ is lepton flavour universal and diagonal \cite{Cirigliano:2009wk,Cata:2015lta}.
A deviation from this prediction would hence
indicate a non-linear realization of electroweak symmetry breaking \cite{Cata:2015lta}. Presently such a deviation is not observed,
however, and we will use  $C_{V_R}^{\ell\ell'}\equiv C_{V_R}\delta_{\ell\ell'}$ in the remainder of this paper.

Another implication of SMEFT is that that the Wilson coefficients $C_{ledq}^{\ell\ell'3i}$, that give rise to
$C_{S_R}^{\ell\ell'}$, also generate neutral-current operators of the form
$(\bar q_L b_R)(\bar \ell_R \ell'_L)$, where $q=d,s,b$. For $q=s$ or $d$, these operators are constrained
very strongly by the leptonic decays $B_{d,s}\to\ell\ell'$, that are forbidden for $\ell\neq\ell'$ and strongly
helicity suppressed for $\ell=\ell'$ in the SM. From existing bounds and measurements, we
find that the SMEFT Wilson coefficients $C_{ledq}^{\ell\ell'31}$ and $C_{ledq}^{\ell\ell'32}$ can induce
effects in $C_{S_R}^{\ell\ell'}$ at most at the per mil level,
which would not lead to any visible effects in $b\to c\ell\nu_{\ell'}$ at the current level of precision.
However, sizable effects in $C_{S_R}^{\ell\ell'}$ cannot be excluded, since the coefficients
$C_{ledq}^{\ell\ell'33}$ contributing to the sum in \eqref{eq:smeftS} only generate the
flavour-conserving operators $(\bar b_L b_R)(\bar \ell_R \ell'_L)$, that
are allowed to be sizable.\footnote{%
Note that these operators do not contribute to
(and thus are not constrained by) leptonic decays of $\Upsilon(nS)$ \cite{Aloni:2017eny}.}

Finally, the availability of the full anomalous dimension matrix for SMEFT dimension-six operators
\cite{Jenkins:2013zja,Jenkins:2013wua,Alonso:2013hga} allows for the prediction of operator mixing due to electroweak
renormalization effects; this will be discussed briefly at the end of section~\ref{sec:models}.

\section{$B\to D^{(*)}$ form factors}\label{sec:ff}

The hadronic form factors of the $B\to D^{(*)}$ transitions are crucial
both for the determination of the CKM element $V_{cb}$ in the
SM and for constraining NP contributions to $b\to c\ell\nu$.
An important difference between the two scenarios is that in the SM $V_{cb}$ only
changes the overall normalization of the rates, but does not modify
the shapes of differential distributions. NP contributions on the other hand can modify these
shapes and can also involve additional form factors,
in particular tensor form factors.
Since our main interest is constraining NP in $b\to c\ell\nu$, we want to use
as much information on the form factors from theory as possible, while at
the same time remaining conservative enough not to introduce ficticious tensions
with the precise experimental data due to too rigid parametrizations.
We therefore use  information from four complementary methods to determine the
$B\to D^{(*)}$ form factors:
\begin{itemize}
\item LQCD. We use all available unquenched LQCD calculations of $B\to D^{(*)}$ form factors.
The $B\to D$ vector and scalar form factors have been computed by the HPQCD \cite{Na:2015kha} and the Fermilab/MILC collaborations
\cite{Lattice:2015rga} at several values of $q^2$, constraining the shape of these form factors in addition to their normalizations.
The FLAG collaboration has performed an average of these two computations, fitted to the BCL parametrization \cite{Aoki:2016frl}.
For the $B\to D^*$ form factors, so far only calculations at zero hadronic recoil
have been reported \cite{Bailey:2014tva,Harrison:2017fmw}; we use their average calculated in \cite{Harrison:2017fmw}.
\item QCD light-cone sum rules allow to compute the form factors in the
region of large hadronic recoil, depending on $B$ meson light-cone distribution amplitudes as non-per\-tur\-bative input
\cite{Faller:2008tr}.
This method is complementary to LQCD, being valid in the opposite kinematic limit. We use the form factor values and
ratios obtained in Ref.~\cite{Faller:2008tr} at\footnote{%
The kinematic variable $w$ is related to $q^2$ as $w= (m_B^2 + m_{D^{(*)}}^2 - q^2) / (2m_Bm_{D^{(*)}})$.}
$q^2=0$, and extract from their values given for $\rho^2_{D,D^*}$ two more, correlated pseudo-datapoints at
$w(q^2)=1.3$.
\item Heavy Quark Symmetry and Heavy Quark Effective Theory (HQET). Treating both the bottom and charm
quark as heavy compared to a typical scale of QCD interactions, QCD exhibits a symmetry among the heavy quarks
\cite{Shifman:1986sm,Isgur:1989vq}. As a consequence, all $B\to D^{(*)}$ form factors either vanish or
reduce to a single function -- the leading Isgur-Wise function -- in the infinite mass limit \cite{Isgur:1989ed}. Perturbative QCD and power
corrections to this limit are partly calculable \cite{Luke:1990eg,Neubert:1991xw,Falk:1992wt,Neubert:1992qq,Neubert:1992wq,
Neubert:1992pn,Ligeti:1993hw,Czarnecki:1996gu,Czarnecki:1997cf,Bernlochner:2017jka}, to be discussed below.
\item By crossing symmetry, the form factors describing the semi-leptonic transition
also describe the pair production of mesons. Unitarity can then be used to impose
constraints on the form factors. Taking into account contributions from other
two-body channels with the right quantum numbers leads via a conservative application of HQET to the \textit{strong
unitarity constraints} \cite{Boyd:1997kz,Caprini:1997mu}. We employ the updated bounds given in \cite{Bigi:2016mdz} for
$B\to D$ and the simplified bounds derived in \cite{Bigi:2017jbd} for $B\to D^*$.
\end{itemize}

We proceed by using the HQET parametrization of all $B\to D^{(*)}$ form factors,
including corrections of $O(\alpha_s, \Lambda_\text{QCD}/m_{c,b})$,
in the notation of \cite{Bernlochner:2017jka}, with two differences:
\begin{enumerate}
\item Instead of using the CLN relation between slope and curvature of the leading Isgur-Wise function, we include both as
independent parameters in the fit.
\item The treatment of higher-order corrections of $\mathcal O(\Lambda_{\mathrm{QCD}}^2/m_c^2)$ has recently been shown to have a
significant effect on the extraction of $V_{cb}$ in the SM, see the discussions in
\cite{Bigi:2016mdz,Aoki:2016frl,Grinstein:2017nlq,Bernlochner:2017jka,Bigi:2017jbd,Bigi:2017njr,Jaiswal:2017rve,Bernlochner:2017xyx}.
We make these corrections explicit by including in addition to the subleading Isgur-Wise functions at order
$\Lambda_{\mathrm{QCD}}/m_{c,b}$ corrections of order $\Lambda_\text{QCD}^2/m_{c}^2$ in those HQET form factors
that are protected from $O(\Lambda_\text{QCD}/m_{c})$ corrections.\footnote{Note that two of these corrections are implicitly included
in \cite{Bernlochner:2017jka} when the normalizations of the $B\to D^{(*)}$ form factors $h_{A_1,+}$ are decoupled from their HQET
values.} From comparison of the HQET predictions at $\mathcal{O}(\alpha_s,\Lambda_{\mathrm{QCD}}/m_{b,c})$, using the three-point sum
rule results for the subleading Isgur-Wise functions \cite{Neubert:1992wq,Neubert:1992pn,Ligeti:1993hw} with LQCD results, we observe
a shift of about $-10\%$ from these corrections in $h_{A_1}(1)$, while the corrections in $h_+(1)$ are only a few per cent. For
$h_{T_1}(1)$ we allow for an independent correction of $10\%$ which is not constrained by lattice data. Note that these corrections are
in principle obtainable from LQCD in all form factors, however, so far only results for those appearing in the SM are available.
\end{enumerate}
We then perform Bayesian and frequentist fits of this parametrization to pseudo data points corresponding
to the described inputs.
The result is a posterior probability distribution or profile likelihood for all the form factor
parameters, respectively.
These can be interpreted as theory predictions for all
$B\to D^{(*)}$ form factors in the entire (semi-leptonic) kinematic range.
This theory prediction is then used in our numerical analysis as a prior on
(in Bayesian fits) or a pseudo-measurement of (in frequentist fits) the
form factor parameters.

\section{$V_{cb}$ from exclusive decays in the Standard Model}\label{sec:Vcb}

Fits for the CKM element $V_{cb}$ from the exclusive decays $B\to D\ell\nu$
and $B\to D^*\ell\nu$ measured at the $B$ factories BaBar and Belle have
already been performed in the literature
(see \emph{e.g.}\
\cite{Bigi:2016mdz,Bigi:2017njr,Bigi:2017jbd,Aoki:2016frl,Lattice:2015rga,Bernlochner:2017jka,Grinstein:2017nlq,Bernlochner:2017xyx,Jaiswal:2017rve}
for recent fits).
Here, we repeat this exercise to
define our assumptions on form factors and our experimental input.
Furthermore, all of our fits are reproducible using open source code,
allowing them to be adapted or modified with different theoretical or experimental
inputs.

We perform fits
to $B\to D\ell\nu$ and $B\to D^*\ell\nu$ decays,
where the theoretical uncertainties are dominated by the hadronic form factors.
To study the impact of different statistical treatments of these ``nuisance''
parameters, we consider three different fits:
\begin{itemize}
\item A frequentist fit where the theoretical knowledge of form factors is treated
as a pseudo-measurement and the individual parameters are profiled over.
\item A Bayesian fit where the theoretical knowledge of form factors is treated
as a prior probability distribution and the individual parameters are marginalized
over.
\item A ``fast fit'' where the theoretical uncertainty on each bin is determined
by varying the form factor parameters according to the theoretical constraints
and is added in quadrature with the experimental uncertainties.
\end{itemize}

\begin{table}[tbp]
\renewcommand{\arraystretch}{1.3}
\centering
\begin{tabular}{lllll}
\hline
Decay & Observable & Experiment & Comment & Ref. \\
\hline
$B\to D(e,\mu)\nu$ & BR & BaBar & global fit & \cite{Aubert:2008yv} \\
$B\to D\ell\nu$ & $d\Gamma/dw$ & BaBar & hadronic tag & \cite{Aubert:2009ac} \\
$B\to D(e,\mu)\nu$ & $d\Gamma/dw$ & Belle & hadronic tag  & \cite{Glattauer:2015teq} \\
\hline
$B\to D^*(e,\mu)\nu$ & BR & BaBar & global fit & \cite{Aubert:2008yv} \\
$B\to D^*\ell\nu$ & BR & BaBar & hadronic tag  & \cite{Aubert:2007qw} \\
$B\to D^*\ell\nu$ & BR & BaBar & untagged $B^0$ & \cite{Aubert:2007rs} \\
$B\to D^*\ell\nu$ & BR & BaBar & untagged $B^\pm$& \cite{Aubert:2007qs} \\
$B\to D^*(e,\mu)\nu$ & $d\Gamma_{L,T}/dw$ & Belle & untagged & \cite{Dungel:2010uk}\\
$B\to D^*\ell\nu$  & $d\Gamma/d(w,\cos\theta_V, \cos\theta_l, \phi)$& Belle & hadronic tag  & \cite{Abdesselam:2017kjf}\\
\hline
\end{tabular}
\caption{Experimental analyses considered. The analyses labeled by $B\to D^{(*)}\ell\nu$ do not differentiate between the lepton
species and are hence not used when analyzing scenarios with non-universal coefficients.}
\label{tab:data}
\end{table}

\subsection{$V_{cb}$ from $B\to D\ell\nu$}\label{sec:VcbD}

The BaBar collaboration has measured the differential branching ratio of
$B\to D\ell\nu$, reconstructed with hadronic tagging \cite{Aubert:2009ac},
in ten bins, averaged over electrons and muons as well as charged and neutral
$B$ decays. Since only statistical uncertainties are given, we follow
\cite{Bigi:2016mdz} and add a fully correlated systematic uncertainty of
$6.7\%$ on the rate. To avoid a bias towards lower values of $V_{cb}$
caused by underfluctuations in individual bins (``d'Agostini bias'' \cite{DAgostini:1993arp}),
we always treat relative systematic errors as relative to the SM predictions
rather than the experimental central values.
In addition to this differential measurement, we also include a BaBar measurement
of the total branching ratio from a global fit, split by electrons and muons
\cite{Aubert:2008yv}, that is statistically independent of the former; following Ref.~\cite{Bigi:2017jbd}, we assume the measurement of
the total branching ratio to be unaffected by the form factor parametrization.
We take into account the significant correlation with the $D^*$
modes extracted in the same analysis.
Following HFLAV \cite{Amhis:2016xyh}, we apply a rescaling of
$-3.7\,\%$ to the published branching ratio to account for updated
$D$ branching ratios.

We note that we cannot use the global HFLAV {\em average} of the $B\to D\ell\nu$ branching ratio because
\begin{itemize}
  \item it contains older measurements that assumed a particular form factor parametrization and would be inconsistent to use in a NP analysis,
  \item it contains measurements from the anaylses that we include in binned form, such that including it would amount to double-counting,
  \item it only considers the average of the electronic and muonic branching ratios, so we cannot use it for the lepton-flavour dependent NP analysis.
\end{itemize}
The Belle collaboration has measured the differential branching ratios separately for electrons and muons as well as charged and
neutral $B$ decays in ten bins each \cite{Glattauer:2015teq}, specifying the full correlation matrix. This measurement does not rely
on a specific form factor parametrization.

\begin{figure}[tbp]
\includegraphics[width=\textwidth]{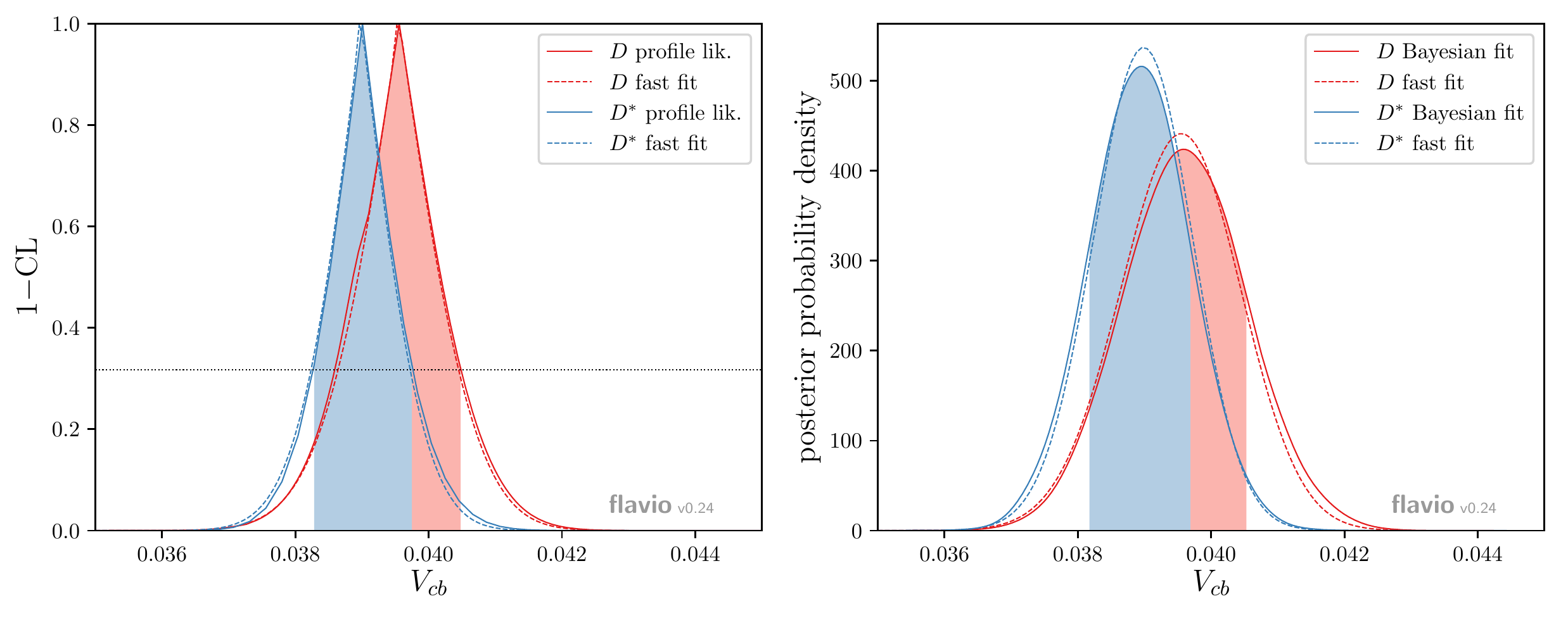}
\caption{Fit results for $V_{cb}$ from exclusive decays in the Standard Model. Left: frequentist profile likelihood
vs.\ ``fast fit'' method; the horizontal dotted line delineates the $1\sigma$ region.
Right: Bayesian posterior probability vs.\ ``fast fit'' method.}
\label{fig:Vcb}
\end{figure}

The combined fit to Belle and BaBar data
with the three different statistical approaches,
using the form factors described
in section~\ref{sec:ff},
yields
\begin{align}
V_{cb}^{B\to D\ell\nu} &= (3.96\pm 0.09)\times  10^{-2} && \text{(frequentist fit)},
\label{eq:VcbD}\\
V_{cb}^{B\to D\ell\nu} &= (3.96\pm 0.09)\times  10^{-2} && \text{(Bayesian fit)},
\\
V_{cb}^{B\to D\ell\nu} &= (3.96\pm 0.09)\times  10^{-2} && \text{(fast fit)},
\end{align}
where the errors in the frequentist fit refer to a change by
$0.5$ in the profile likelihood and in the Bayesian case correspond to the highest
posterior density interval with $68.3\%$ credibility.
The agreement of the central value between the frequentist and the Bayesian fit
is not surprising as the minimum of the two likelihoods coincides (the prior
probability for theory parameters has the same mathematical form as the
``pseudo measurements'' used in the frequentist fit). That the errors from
the profile likelihood and the posterior marginalization agree is less trivial,
but is a consequence of all theory and experimental uncertainties being close
to Gaussian.
The excellent agreement of the two more sophisticated approaches with the
``fast fit'' approach
indicates that the fitted
values of the nuisance parameters are close to the theoretical central values
(otherwise a mismatch would be observed in the fast fit).
The full one-dimensional profile likelihood and posterior probability distribution
are shown in figure~\ref{fig:Vcb}.

We observe good compatibility with other recent extractions from this mode
\cite{Bigi:2016mdz,Bernlochner:2017jka,Jaiswal:2017rve}. A direct comparison is not possible, since neither the form factor
parametrizations nor the data sets used are identical. The inclusion of additional data in our case is responsible for the slightly
smaller uncertainties. Note that the inclusion of the measurement \cite{Aubert:2008yv} shifts $V_{cb}$ to smaller values compared to
\cite{Bigi:2016mdz}.

In our numerical analysis of new physics effects in section~\ref{sec:np},
when allowing for lepton flavour non-universal effects, we only use measurements
where electron and muon samples are separated, since the generally different but unknown
electron and muon efficiencies prohibit a consistent interpretation of the
combined measurements in such scenarios. It is instructive to extract the
value of $V_{cb}$ only from these subsets of measurements. Using the frequentist
approach, we find
\begin{align}
V_{cb}^{B\to De\nu} &= (4.00^{+0.07}_{-0.17})\times  10^{-2}
\,,\\
V_{cb}^{B\to D\mu\nu} &= (3.96^{+0.11}_{-0.10})\times  10^{-2}
\,.
\end{align}
We observe consistency among these values and with the global fit, albeit with larger uncertainties.

\subsection{$V_{cb}$ from $B\to D^*\ell\nu$}

Since the $D^*$ is a vector meson and further decays to \emph{e.g.}\ $D\pi$, a four-fold
differential decay distribution in three angles and $w$
can be
measured. Belle has recently published an analysis with one-dimensional
distributions in all four kinematic quantities with full error correlations,
based on hadronic tagging
\cite{Abdesselam:2017kjf}.
An earlier (and statistically independent) untagged analysis by Belle
\cite{Dungel:2010uk} considers the $w$-differential branching ratios into
longitudinally or transversely polarized $D^*$, that can be extracted from the
angular distribution.
Since the error correlations are not publicly available, we simply assume the statistical
uncertainties to be fully uncorrelated and the systematic ones to be fully
correlated. In addition, we rescale the systematic uncertainties, that are
given as relative uncertainties with respect to the measured central values,
into relative uncertainties with respect to the SM prediction instead, again to avoid
the D'Agostini bias mentioned before.
In addition to the differential measurements, we include four
measurements of the total branching ratio by BaBar,
listed in table~\ref{tab:data}, which are statistically independent
to a good approximation~\cite{Hamano:2008iba}.
As mentioned in section~\ref{sec:VcbD}, we do however take into account
the significant correlation between the $B\to D^*(e,\mu)\nu$
and $B\to D(e,\mu)\nu$ measurements of the BaBar ``global fit''.
As in the case of $B\to D\ell\nu$, we cannot use the HFLAV average for the $B\to D^*\ell\nu$ branching ratio, but we
apply the same rescalings as HFLAV to account e.g.\ for changes
in $D^*$ branching ratios. The central value of the BaBar
global fit is hardly modified, but the other three branching ratio measurements are reduced by 5--6\% compared to the
published values.

The combined fit to these data,
using our form factor parametrization with the constraints discussed in Sec.~\ref{sec:ff},
yields
\begin{align}
V_{cb}^{B\to D^*\ell\nu} &= (3.90\pm 0.07)\times  10^{-2} && \text{(frequentist fit)},
\\
V_{cb}^{B\to D^*\ell\nu} &= (3.90\pm 0.07)\times  10^{-2} && \text{(Bayesian fit)},
\\
V_{cb}^{B\to D^*\ell\nu} &= (3.90\pm 0.07)\times  10^{-2} && \text{(fast fit)}.
\end{align}
Again, we observe excellent agreement of the three different approaches
and the same comments as in section~\ref{sec:VcbD} apply.
We conclude that the extraction of $V_{cb}$ from exclusive decays does not
depend in a relevant way on the statistical approach taken.
The extracted values are also comparable to those in the recent literature.

Again we observe good compatibility with other recent extractions from this mode
\cite{Bigi:2017njr,Bernlochner:2017jka,Grinstein:2017nlq,Jaiswal:2017rve}, and the same comments regarding comparability apply as in
$B\to D$. The inclusion of additional data in our case is responsible for the slightly smaller uncertainty and lower central value. The
latter is also related to a shift from using the experimental values from HEPData, as discussed below.

As for $B\to D\ell\nu$ at the end of section~\ref{sec:VcbD},
we also repeat the extraction of $V_{cb}$ using only the measurements
that separate the electron and muon samples, since these measurements are
used in lepton flavour non-universal NP scenarios in section~\ref{sec:np}.
Using the frequentist
approach, we find
\begin{align}
V_{cb}^{B\to D^*e\nu} &= (3.89\pm0.10)\times  10^{-2}
\,,\\
V_{cb}^{B\to D^*\mu\nu} &= (3.76\pm0.11)\times  10^{-2}
\,.
\end{align}
While consistent with each other and with the global fit, we observe that
the muonic data prefer a value of $V_{cb}$ that is lower by about
one standard deviation.

Finally we would like to comment on the robustness of the extracted $V_{cb}$ value from binned data.
The value obtained from the Belle 2017 data alone depends on the precise inputs used: the difference
between the binned data given on HEPData \cite{1512299} and that in the publication, where the values are rounded to two significant digits,
yields a shift in $V_{cb}$ of about one standard deviation.\footnote{This might also explain the difference between the values of
$V_{cb}$ obtained from unfolded and folded data in this article.} This is problematic, since the uncertainties and correlations
themselves have an uncertainty that is likely to be larger than the difference between these inputs.
This should be kept in mind when analyzing the unfolded spectrum.
We proceed using the inputs given on HEPData where available.

\section{New physics}\label{sec:np}

Having extracted the CKM element $V_{cb}$ from data assuming the
SM, we next proceed to constrain NP effects. As discussed in section~\ref{sec:heff},
there are five Wilson coefficients per lepton-flavour combination, with relations
for the right-handed vector current $C_{V_R}^{\ell\ell'}\equiv C_{V_R}\delta_{\ell\ell'}$, \emph{i.e.} up to 25 independent
Wilson coefficients
for $b\to c(e,\mu)\nu$ transitions. Since the ones for $\ell\neq \ell'$ are indistinguishable, these are
effectively reduced to 17, and in the lepton-flavour diagonal case only 9 operators remain.
Before analyzing them in more detail, we discuss in section~\ref{sec:models} all possible
simplified models that can be generated by the exchange of a single new mediator particle,
implying specific combinations of these Wilson coefficients.

While we focus on exclusive modes in the SM $V_{cb}$ fits, for the NP fits
we also compare to the inclusive decays $B\to X_c \ell\nu$. Since the full inclusive
analysis involves fitting several moments of the spectrum simultaneously with
$V_{cb}$, quark masses, and HQET parameters, reproducing it is beyond the scope
of our present analysis. Rather, we use a simplified approach where we
approximate the total rate in the presence of new physics as
\begin{equation}
\Gamma(B\to X_ce\nu) \approx \Gamma(B\to X_ce\nu)_\text{SM}
~
\frac{\Gamma(B\to X_ce\nu)_\text{NP}^\text{LO}}{\Gamma(B\to X_ce\nu)_\text{SM}^\text{LO}}\,,
\end{equation}
where $\Gamma(B\to X_ce\nu)_\text{SM}$ is the full state-of-the-art
SM prediction \cite{Alberti:2014yda} and the rates with superscript ``LO'' refer to the
partonic leading-order calculations.
We confirm the known expressions for the LO contributions in the $m_\ell\to0$ limit
\cite{Grossman:1994ax,Dassinger:2007pj,Colangelo:2016ymy,Celis:2016azn}:
\begin{align}
\Gamma^{\text{LO}}(B\to X_c\ell\nu) = &\,\Gamma^{\text{LO}}_{\text{SM}}(B\to
X_c\ell\nu)\left[\left(|1+C_{V_L}^\ell|^2+|C_{V_R}|^2\right)+\frac{1}{4}\left(|C_{S_L}^\ell|^2+|C_{S_R}^\ell|^2\right)
+12|C_T^\ell|^2\right]\nonumber\\
&\,+\Gamma_\text{Int}^\text{LO}(B\to X_c\ell\nu)\left[{\text{Re}}\left[(1+C_{V_L}^\ell)
C_{V_R}^*\right]-\frac{1}{2}{\text{Re}}(C_{S_L}^\ell C_{S_R}^{\ell\,*})\right]\,,\quad\mbox{with}\\
\Gamma^{\text{LO}}_{\text{SM}}(B\to X_c\ell\nu) = & \,\Gamma_{0}\left(1-8x_c-12x_c^2\log x_c+8x_c^3-x_c^4\right)\quad\mbox{and}\\
\Gamma_\text{Int}^\text{LO}(B\to X_c\ell\nu) = &\, -4\Gamma_0 \sqrt{x_c}\left(1+9x_c+6x_c(1+x_c)\log x_c-9x_c^2-x_c^3\right)\,,
\end{align}
and $\Gamma_0=G_F^2 m_b^5|V_{cb}|^2/(192\pi^3)$. $1/m_{c,b}^2$ and $\alpha_s$ corrections to the NP contributions are partly known
and can be sizable
\cite{Jezabek:1988iv,Czarnecki:1992zm,Czarnecki:1994bn,Grossman:1995yp,Dassinger:2007pj,Colangelo:2016ymy,Celis:2016azn}. Specifically
for scalar operators the $\alpha_s$ corrections can qualitatively change the result \cite{Celis:2016azn}. However, since we do not include effects
on the spectra in any case, we stick to the simple expressions given above; we therefore do not consider the inclusive constraints used
here as on the same footing as the exclusive ones and do not combine them in a global fit.

Finally, we include the constraint from the total width of the $B_c$ meson, which can be modified significantly in scenarios with
scalar couplings \cite{Li:2016vvp,Alonso:2016oyd}.

\subsection{Tree-level models}\label{sec:models}

Since the $b\to c\ell\nu$ transition is a tree-level process in the SM,
NP models with sizable effects in these modes typically involve tree-level
contributions as well.
The known quantum numbers of the involved fermions allow to determine all possible mediator quantum numbers.
These correspond to the following simplified models:
\begin{itemize}
 \item New vector-like quarks modifiying the $W$-couplings of the SM quarks,
 \item tree-level exchange of a heavy charged vector boson ($W'$),
 \item tree-level exchange of a heavy charged scalar ($H^\pm$),
 \item tree-level exchange of a coloured vector or scalar boson (leptoquark).
\end{itemize}

Vector-like quarks are
among the simplest renormalizable extensions of the SM. They are
also present, \emph{e.g.}, in models with partial compositeness in
the form of fermionic resonances of the strongly interacting sector.
An $SU(2)_L$ singlet vector-like quark with mass $m_\Psi$, coupling to the SM
quark doublet and the Higgs with strength $Y$, can generate a modified
left-handed $W$ coupling of order $Y^2v^2/m_\Psi^2$, while an $SU(2)_L$ doublet,
coupling to the SM quark singlet and the Higgs, can generate a right-handed
$W$ coupling. The resulting contributions to the Wilson coefficients
$C_{V_L}$ or $C_{V_R}$, respectively, are lepton-flavour universal.

The tree-level exchange of $W'$, $H^\pm$, or leptoquarks has been studied extensively
in the literature in the context of explanations of the $b\to c\tau\nu$
anomalies
(see e.g.\ \cite{
Crivellin:2012ye,Celis:2012dk,Tanaka:2012nw,Greljo:2015mma,
Freytsis:2015qca,
Calibbi:2015kma,Fajfer:2015ycq,Barbieri:2015yvd,Bauer:2015knc,Das:2016vkr,Becirevic:2016yqi
}.)
The case of leptoquarks is particularly diverse as there are six different
representations -- three scalars and three vectors -- that can in principle
contribute to $b \to c\ell\nu$ at tree level. The $b \to c\ell\nu$ Wilson
coefficients generated in each of the tree-level models considered are shown
in table~\ref{tab:models}. Crosses correspond to a possible tree-level effect at the
matching scale, corresponding to the scale of new physics.
If two crosses appear in a line, the contributions are governed by independent
parameters in the model.

An interesting observation concerning table~\ref{tab:models} is that none
of the models generates the tensor operator $O_T$ on its own. At the matching
scale, it is only present in the two scalar leptoquark models $S_1$ and $R_2$
in a characteristic correlation with the operator $O_{S_L}$.

\begin{table}
\centering
\renewcommand{\arraystretch}{1.2}
\begin{tabular}{cccccccc}
\hline
Model & $C_{V_L}$ & $C_{V_R}$ & $C_{S_R}$ & $C_{S_L}$ & $C_T$ & $C_{S_L}=4C_T$ & $C_{S_L}=-4C_T$\\
\hline
Vector-like singlet & $\times$ & \\
Vector-like doublet && $\times$ & \\
$W'$ & $\times$ & \\
$H^\pm$ &&& $\times$ & $\times$ & \\
$S_1$ & $\times$ &&&&&& $\times$ \\
$R_2$ &&&&&& $\times$ \\
$S_3$ & $\times$ \\
$U_1$ & $\times$ && $\times$ \\
$V_2$ &&& $\times$ \\
$U_3$ & $\times$ \\
\hline
\end{tabular}
\caption{Pattern of Wilson coefficients generated by tree-level models.
Crosses correspond to a possible effect at the matching scale.
If two crosses appear in a line, the contributions are governed by independent
parameters in the model.}
\label{tab:models}
\end{table}

Since table~\ref{tab:models} is valid at the new physics scale,
it is also important to consider renormalization group (RG) effects between
this scale and the $b$ quark scale that could possibly change this picture.
Normalized as in Eq.~\eqref{eq:ops}, it can be easily seen that the vector
operators are not renormalized under QCD as they correspond to
conserved currents;
the same is true for the combinations $m_b O_{S_{L,R}}$, such that
the Wilson coefficients renormalize under QCD like the quark mass.
Thus, the operator $O_T$, which is mildly QCD-renormalized,
cannot mix with the other operators under QCD.
Therefore the only qualitatively relevant RG mixing effects among the five operators of
interest could come from electroweak
renormalization in the SMEFT above the electroweak scale.
Indeed, the operators $O_{lequ}^{(1)}$ and $O_{lequ}^{(3)}$, that match
onto $O_{S_L}$ and $O_T$, respectively, mix with each other through the
weak coupling constants \cite{Gonzalez-Alonso:2017iyc}. The RG-induced values of the respective coefficients
at the electroweak scale $v$ can be written in leading logarithmic approximation
as
\begin{align}
 C_{S_L}(v) \supset
 \frac{3}{8\pi^2}
 (3g^2+5g^{\prime2})
 \,\ln\!\left(\frac{v}{\Lambda_\text{NP}}\right)
 \, C_T(\Lambda_\text{NP})\,,
\\
 C_T(v) \supset
 \frac{1}{128\pi^2}
 (3g^2+5g^{\prime2})
 \,\ln\!\left(\frac{v}{\Lambda_\text{NP}}\right)
 \, C_{S_L}(\Lambda_\text{NP})\,,
\end{align}
where the self-mixing contributions have been omitted for brevity, but are included in the numerical analysis.
For $\Lambda_\text{NP}=1\,\text{TeV}$,
the numerical prefactors are roughly $-0.1$ and $-0.002$, respectively,
so the effect is small unless the scale separation is very large,
in particular given that the tree-level models already predict $C_{S_L}\gg C_T$
at the matching scale.

We thus conclude that table~\ref{tab:models} is useful as classification
of tree-level models in terms of low-energy effects and is qualitatively
stable under quantum corrections. Of course, a realistic model may combine
several of the discussed particles and thus lead to a more diverse
pattern of effects.

\subsection{Right-handed currents}

$C_{V_R}$ has been considered as a possible source for a tension between inclusive and exclusive determinations in $V_{cb}$ for a long
time \cite{Voloshin:1997zi}.
However, updated analyses based on total rates alone have already shown that the
scenario is disfavoured as a solution to the present tension \cite{Crivellin:2014zpa}.

\begin{figure}[tbp]
\centering
\includegraphics[width=0.5\textwidth]{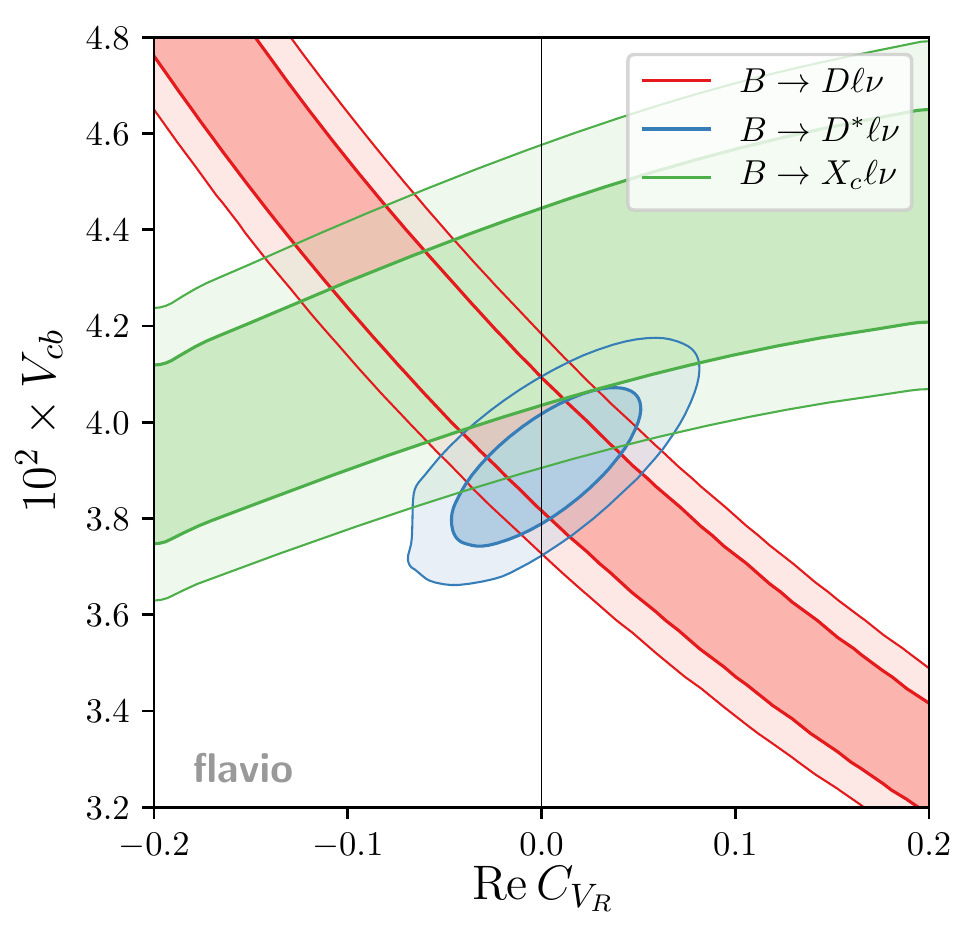}
\caption{Constraints on right-handed currents from inclusive and exclusive
decays, assuming LFU.}
\label{fig:CV}
\end{figure}

The main novelty of the present analysis is the inclusion of the
$B\to D^*\ell\nu$ angular observables in the fit. For the total rate, an effect
in $C_{V_R}$ can always be absorbed by an appropriate shift in $V_{cb}$
(or $\tilde V_{cb}$), such that constraining right-handed currents requires
considering several modes. Including angular observables changes this picture,
as the shape of these observables is directly sensitive to right-handed currents.
As a consequence, a fit to $B\to D^*\ell\nu$ data alone is able to
constrain right-handed currents directly, as shown\footnote{%
This plot and all the following Wilson coefficient plots correspond to the two-dimensional
profile likelihood in the space of the Wilson coefficients shown.
For observables only constraining a single combination of Wilson coefficients,
the bands correspond to $-2\Delta\ln L=1$ and 4, respectively,
otherwise to $-2\Delta\ln L=2.30$ and $6.18$, thereby accounting for the different degrees of freedom.}
in figure~\ref{fig:CV}, together with the constraints from $B\to D\ell\nu$ and the inclusive decay, both constraining only combinations of
$V_{cb}$ and $C_{V_R}$. In this plot we assume lepton-flavour universality as discussed above.

We observe that the tensions between the $V_{cb}$ determinations from
$B\to D\ell\nu$, $B\to D^*\ell\nu$, and $B\to X_c\ell\nu$,
that are anyway milder than in the past,
cannot be completely removed by postulating new physics in right-handed currents.
What is new is that even $B\to D^*\ell\nu$ alone cannot be brought into perfect
agreement with
$B\to X_c\ell\nu$ for any value of $C_{V_R}$.

\subsection{Lepton flavour universality violation}

In view of the observed tensions with SM expectations in $b\to c\tau\nu$ and $b\to s\ell\ell$
transitions, investigating $e$-$\mu$ universality in $b\to c\ell\nu$ with light
leptons is important. Specific new physics models suggested as solutions
to the $b\to c\tau\nu$ anomalies actually predict such violation. Some of
the experimental analyses \textit{assume} LFU to hold.
These analyses cannot be used in a model-independent fit allowing for LFU violation.
This is because the measurements are not simply averages of the respective electron
and muon observables, but linear combinations with weights depending on the
experimental efficiencies that can differ between electrons and muons even
as a function of kinematical variables. Thus it is of paramount importance
that experimental collaborations present their results separately for electrons and muons.

\begin{figure}[tbp]
\centering
\includegraphics[width=0.5\textwidth]{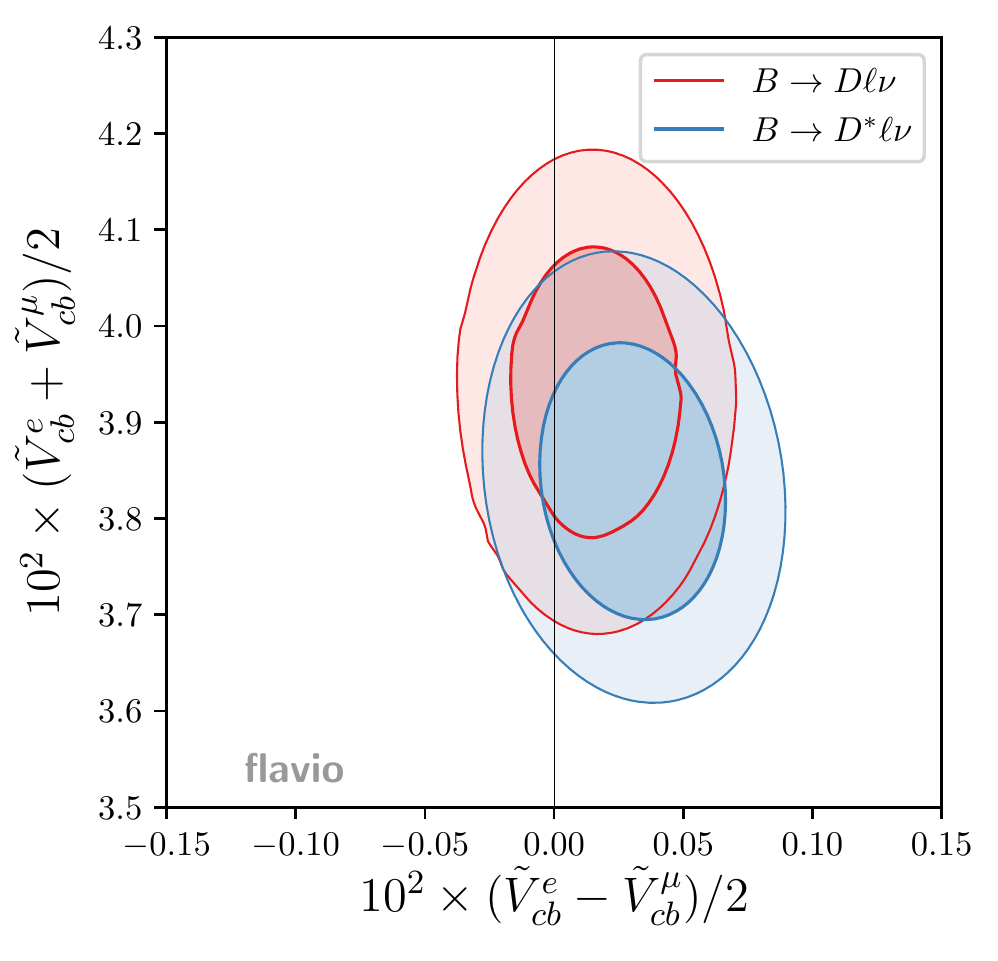}
\caption{Constraints on LFU violation in the left-handed vector current from exclusive decays.
The vertical line corresponds to the LFU limit.}
\label{fig:LFUV}
\end{figure}

In the meantime, the existing analyses that already include separate results
for electrons and muons (see table~\ref{tab:data}) can be used to perform
a fit with a non-universal modification of the SM operator, i.e.\
$C_{V_L}^e\neq C_{V_L}^\mu$. The fit result in terms of the lepton-flavour-dependent
effective CKM elements $\tilde V_{cb}^\ell$
is shown in figure~\ref{fig:LFUV}.
Both for $B\to D\ell\nu$ and $B\to D^*\ell\nu$ the fit not only shows perfect
agreement with LFU, but also implies a stringent constraint on departures from the
LFU limit.
Given the good agrement of the constraints from $B\to D\ell\nu$ and $B\to D^*\ell\nu$, we have also performed a combined Bayesian fit of the scenario to both decay modes, marginalizing over all nuisance parameters. We find
\begin{align}
\frac{1}{2}\left(\tilde V_{cb}^e+\tilde V_{cb}^\mu\right) &= (3.87 \pm 0.09)\%
\,,\\
\frac{1}{2}\left(\tilde V_{cb}^e-\tilde V_{cb}^\mu\right) &= (0.022 \pm 0.023)\%
\,,
\end{align}
with a small correlation of $-10\%$. Equivalently, we find
\begin{equation}
\frac{\tilde V_{cb}^e}{\tilde V_{cb}^\mu} = 1.011 \pm 0.012
\,,
\end{equation}
which can be used as a generic constraint on $e$-$\mu$ universality violation in $b\to c\ell\nu$ processes.
This is the first global combination of LFU constraints in $b\to c\ell\nu$ transitions and provides a significantly stricter bound on
LFU than the recent measurement in Ref.~\cite{Abdesselam:2017kjf} alone.

As already discussed in section~\ref{sec:heff}, a violation of LFU can also
manifest itself as a contribution from a ``wrong-neutrino'' decay
generated by $C_{V_L}^{\ell\ell'}$ with $\ell\neq\ell'$.
However, this case does not have to be discussed separately as it merely leads
to a rescaling of all observables that can be absorbed by defining
\begin{equation}
\tilde V_{cb}^\ell = V_{cb} \bigg[ |1 + C_{V_L}^\ell|^2 + \sum_{\ell'\neq\ell}|C_{V_L}^{\ell\ell'}|^2 \bigg]^{1/2}.
\end{equation}

\subsection{Scalar operators}

The interference of scalar contributions with the SM amplitude in exclusive
and inclusive decays arises only from lepton-mass-suppressed terms. Consequently,
scalar contributions always lead to an \textit{increase} in the rates for a
fixed value of $V_{cb}$. The suppression of interference terms also implies
that there is no qualitative difference
between operators with $\ell=\ell'$ or $\ell\neq\ell'$,
so we focus on the former for simplicity.

\begin{figure}[tbp]
\includegraphics[width=0.55\textwidth]{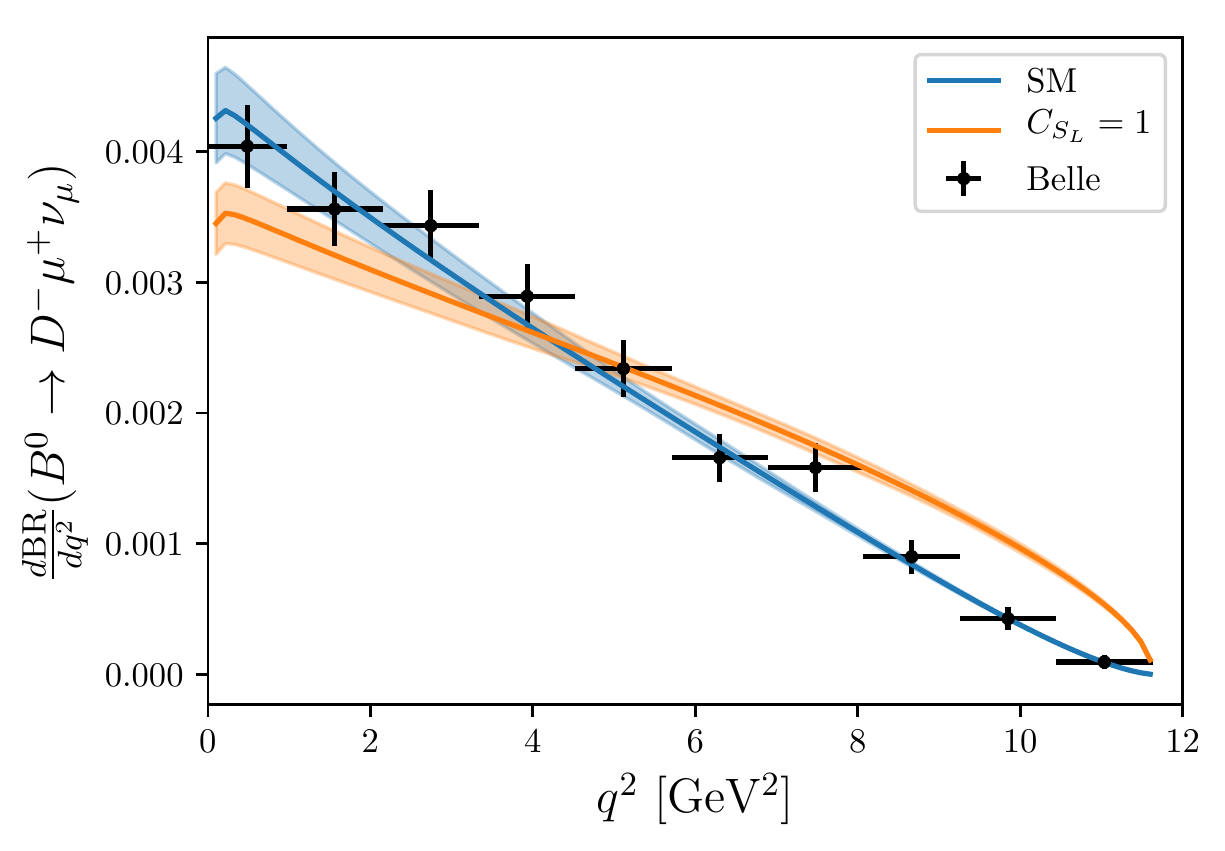}%
\includegraphics[width=0.45\textwidth]{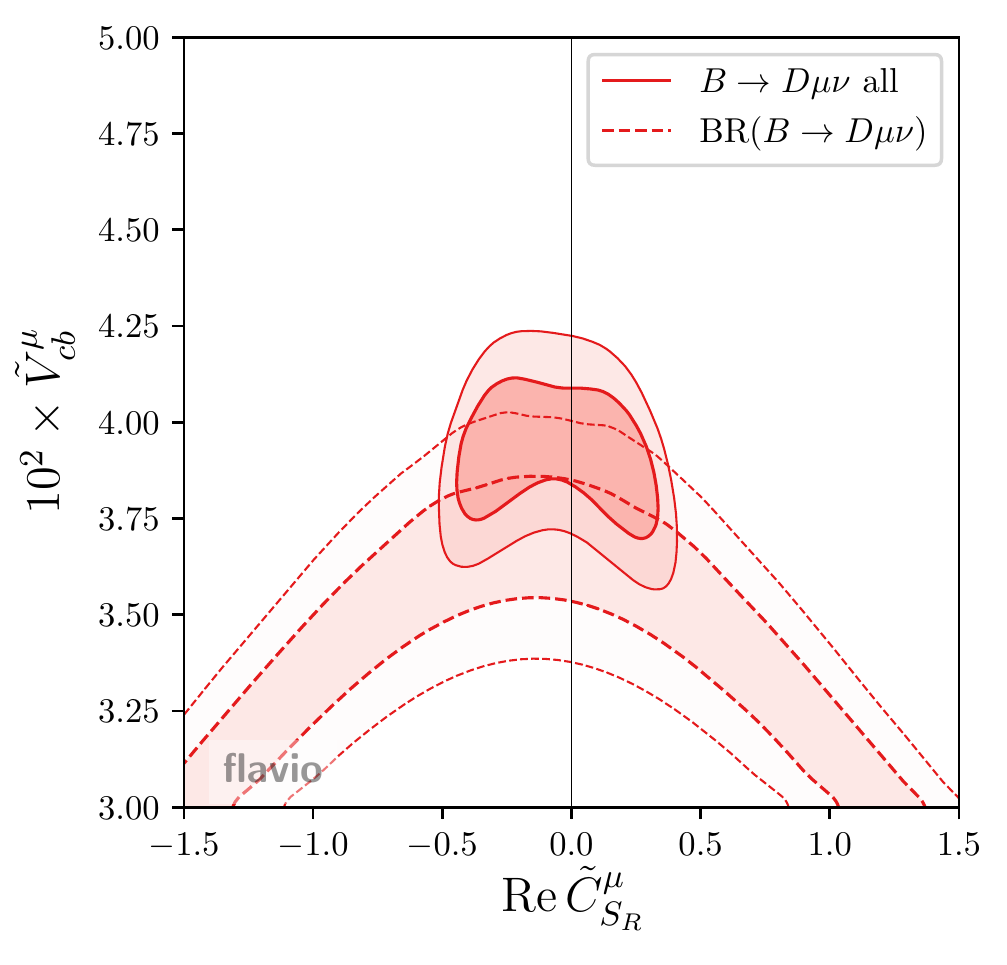}%
\caption{Left: Prediction for the differential $B\to D\mu\nu$ branching ratio
in the SM (blue band) and a scenario with new physics in $C_{S_L}$ (orange
band) vs.\ the Belle measurement, demonstrating the different endpoint
behaviour at zero recoil ($q^2\approx 11.6~\text{GeV}^2$).
Both scenarios predict the same total $B\to D\mu\nu$ branching ratio.
Right: Comparison of the constraint on the scalar coefficient $\tilde C_{S_R}^\mu$ vs.\
$\tilde V_{cb}^{\mu}$ from the total $B\to D\mu\nu$ branching ratio measurements only (dashed) and using all $B\to D\mu\nu$
measurements (solid).}
\label{fig:CSL_q2}
\end{figure}

Again
the differential distributions contain valuable
information about possible scalar contributions.  The most striking example
is the endpoint of the $B\to D\ell\nu$ differential decay rate, see also, \emph{e.g.}, \cite{Nierste:2008qe}. In the SM,
close to $q^2=q^2_\text{max}=(m_B-m_D)^2$, it behaves as
\begin{equation}
\frac{d\Gamma(B\to D\ell\nu)}{dq^2} \propto f_+^2\left(q^2-q^2_\text{max}\right)^{3/2},
\end{equation}
while in presence of NP contributions to scalar operators, it behaves as
\begin{equation}
\frac{d\Gamma(B\to D\ell\nu)}{dq^2} \propto f_0^2|C_{S_R}+C_{S_L}|^2\left(q^2-q^2_\text{max}\right)^{1/2}.
\end{equation}
This implies an exceptional sensitivity of the highest $q^2$ (or lowest $w$)
bin in $B\to D\ell\nu$ to the sum of scalar Wilson coefficients, $C_{S_R}+C_{S_L}$.
To illustrate this effect, in figure~\ref{fig:CSL_q2} we show on the left the predictions
for the SM (using $V_{cb}$ from eq.~\eqref{eq:VcbD}) and a scenario
with a large NP contribution in $C_{S_L}$ for fixed $V_{cb}$, compared to the experimental
data. The scenarios are chosen such that they give the same prediction
for the total branching ratio. The qualitatively different endpoint behaviour
can clearly distinguish between them, excluding such large scalar contributions.

As a consequence, a fit to the differential $B\to D\ell\nu$ rates
leads to a more stringent constraint on scalar operators than
$B\to D^*\ell\nu$ or the inclusive decay.
While the total rates of these modes constrain only combinations of $|V_{cb}|$ and $|C_{S_L}\pm C_{S_R}|$ or $|C_{S_L}|^2+|C_{S_R}|^2$,
these are accessible individually with differential distributions. This is demonstrated in figure~\ref{fig:CSL_q2} on the right, where
the constraint resulting from the total rate alone is compared to the one resulting from all available information on that channel.
Importantly, including the information from the differential rate, $\tilde V_{cb}$ and $\mathrm{Re}(C_{S_R})$ can be
determined individually.

\begin{figure}[tbp]
\centering{
\includegraphics[width=0.5\textwidth]{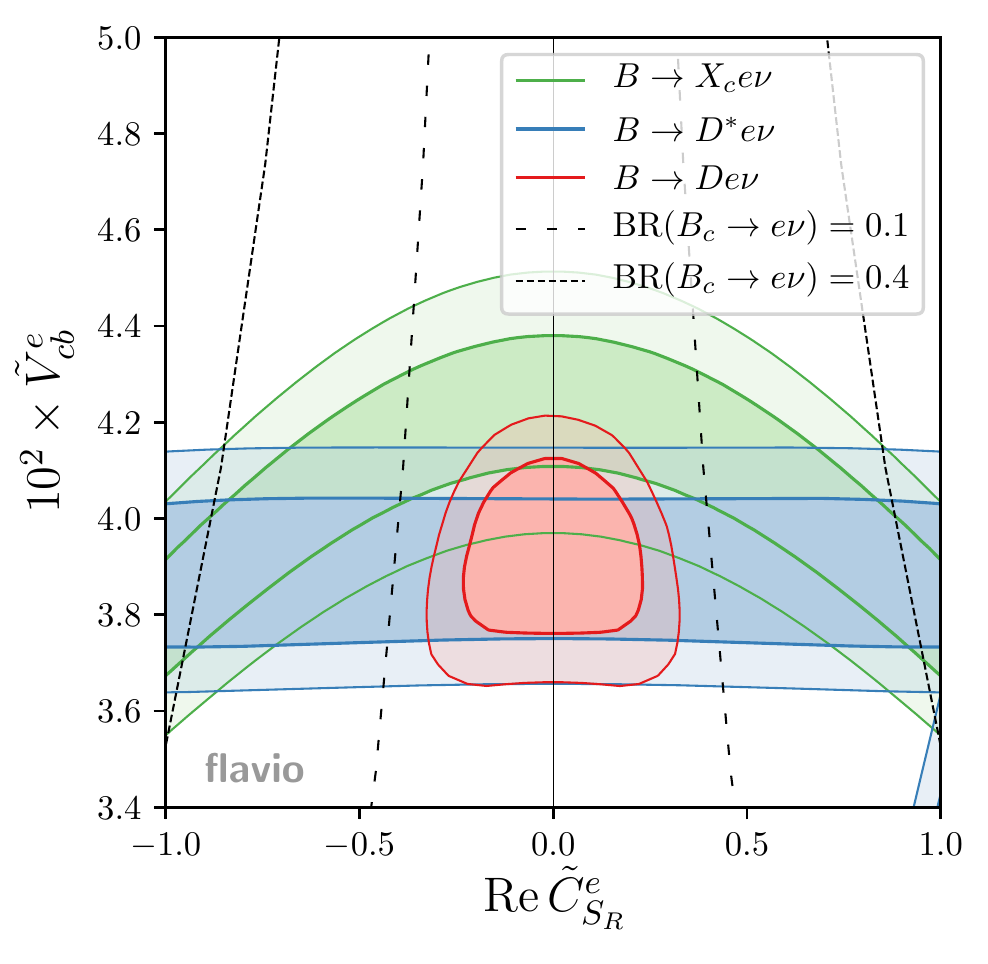}%
\includegraphics[width=0.5\textwidth]{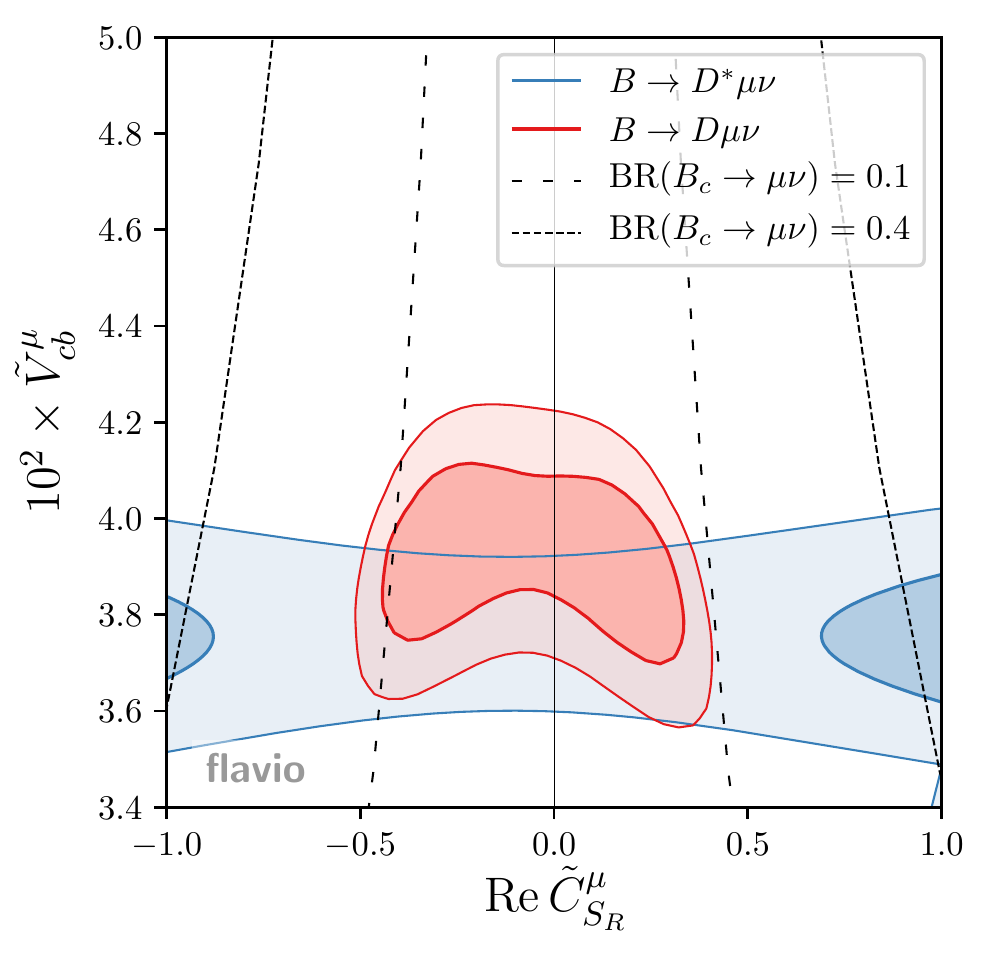}%
}
\caption{Constraints on the scalar coefficients $\tilde C_{S_R}^{e,\mu}$ vs.\
$\tilde V_{cb}^{e,\mu}$, absorbing a potential vector coefficient $C_{V_L}$,
from inclusive and exclusive decays as well as the total width of the $B_c$ meson, seperately for electrons (left) and muons (right).}
\label{fig:CSR}
\end{figure}

\begin{figure}[tbp]
\centering{
\includegraphics[width=0.5\textwidth]{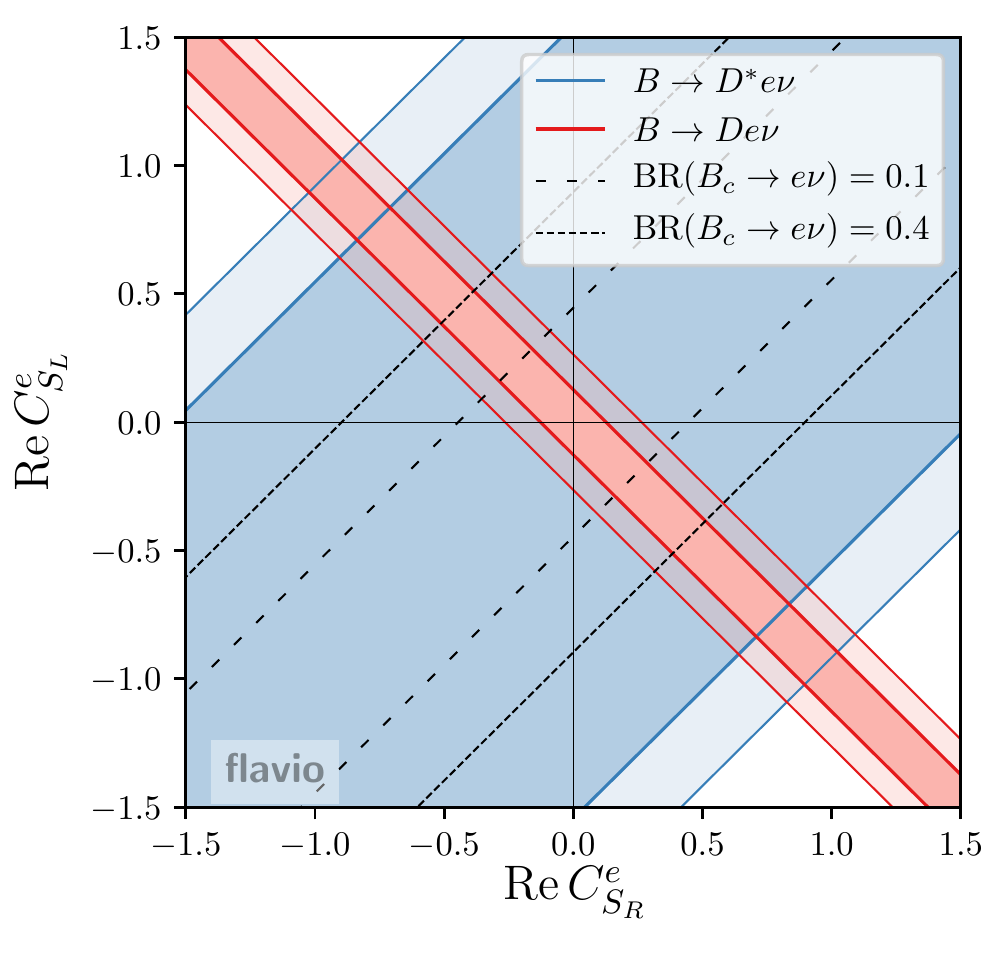}%
\includegraphics[width=0.5\textwidth]{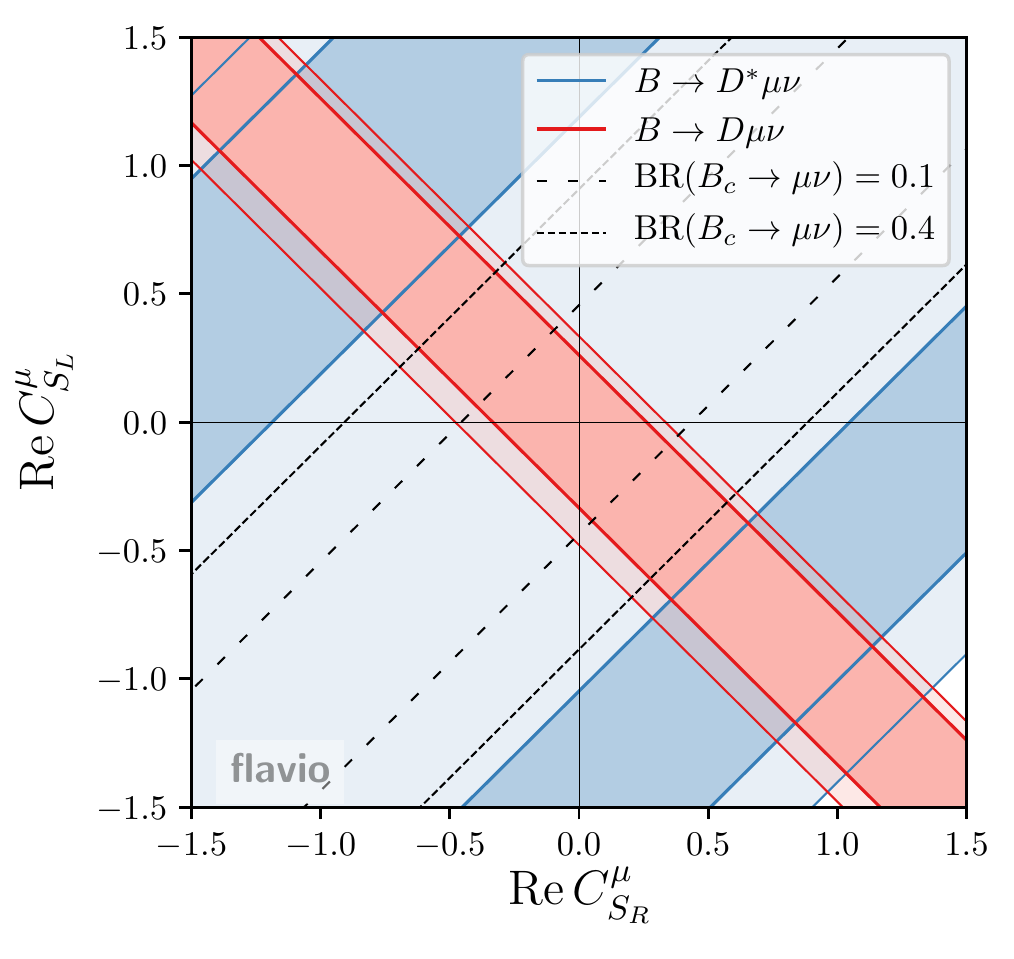}%
}
\caption{Constraints on the scalar Wilson coefficients from exclusive decays and the total width of the $B_c$ meson, seperately for
electrons (left) and muons (right).}
\label{fig:CSR_CSL}
\end{figure}

The constraining power of $B\to D\ell\nu$ and the complementary dependence
of $B\to D\ell\nu$ and $B\to D^*\ell\nu$ on
the scalar Wilson coefficients $C_{S_L}$ and $C_{S_R}$ is also
demonstrated in figures~\ref{fig:CSR} and \ref{fig:CSR_CSL}.
In these figures, the constraints are shown separately for the modes involving electron and muons, as
there is in general no reason to expect LFU to hold for scalar contributions. Furthermore, here and in the following only the
measurements of electronic modes are used to constrain the electronic coefficients and the measurements of muonic modes for the muonic
coefficients. We stress that more stringent constraints would be obtained by
assuming the coefficient with the other lepton flavour to be free from NP
and using both sets of data. In this respect, our constraints are conservative.

In figure~\ref{fig:CSR}, corresponding to scenarios with a $U_1$ or $V_2$
leptoquark, the electronic modes do not show any preference for non-zero $C_{S_R}^e$,
while $B\to D^*\mu\nu$ prefers a sizable value of $C_{S_R}^\mu$ slightly over
the SM, but such values are excluded by $B\to D\mu\nu$ in this scenario, and also disfavoured by the $B_c$ lifetime.

A second interesting scenario is having NP  in both $C_{S_R}$ and  $C_{S_L}$,
as can arise in models with a charged Higgs boson (cf.~table~\ref{tab:models}).
The resulting constraints are shown in figure~\ref{fig:CSR_CSL},
again separately for electronic and muonic modes.
In addition to the form factor parameters, also $V_{cb}$ is varied as a
nuisance parameter. Consequently, the inclusive decay alone, only entering our
analysis via its total rate, cannot constrain the coefficients individually
and is not shown.
In principle, the same qualification applies to the leptonic decays $B_c\to\ell\nu$.
To give an indication, we nevertheless show the constraints that would apply for a conservatively low value of $V_{cb}=0.035$.
The feature that $B\to D\ell\nu$ only constrains the sum
and $B\to D^*\ell\nu$ only the difference of these Wilson coefficients is clearly seen.
In the muonic case, we observe a mild preference for non-zero values
of both Wilson coefficients from $B\to D^{(*)}\ell\nu$. In constrast to the scenario in figure~\ref{fig:CSR},
the constraint from $B\to D\ell\nu$ does not exclude sizable NP effects
if $C_{S_R}\sim C_{S_L}$. However, the constraint from the $B_c$ lifetime excludes such a scenario, even when allowing for very large
modifications of $\sim 40\%$ and a small value of $V_{cb}=0.035$.

\subsection{Tensor operator}

In contrast to the scalar operators, for tensor operators 
$B\to D^*\ell\nu$ leads to a much more stringent constraint. This is due to
the existing lepton-specific data on the transverse and longitudinal decay rates
(cf.\ table~\ref{tab:data}). In the SM the $D^*$ is fully longitudinally polarized at maximum recoil ($q_\text{min}^2=m_\ell^2$).
This is no longer true in the presence of tensor operators.
Indeed, in the limit of $q^2\to q_\text{min}^2\approx 0$, the differential rate to
transversely polarized $D^*$ in the presence of NP contributions to
$C_{V_L}$ and $C_T$ behaves as
\begin{equation}
\frac{d\Gamma_T(B\to D^*\ell\nu)}{dq^2} \propto
q^2 \,|1+C_{V_L}|^2\left( A_1(0)^2 + V(0)^2 \right)
+ 16m_B^2 \,|C_T|^2 \,T_1(0)^2 + O\left(\frac{m_{D^*}^2}{m_B^2}\right),
\end{equation}
where the kinematic relation $T_2(0)=T_1(0)$ was used and the $D^*$ mass was
neglected for simplicity.
\begin{figure}[tbp]
\includegraphics[width=0.55\textwidth]{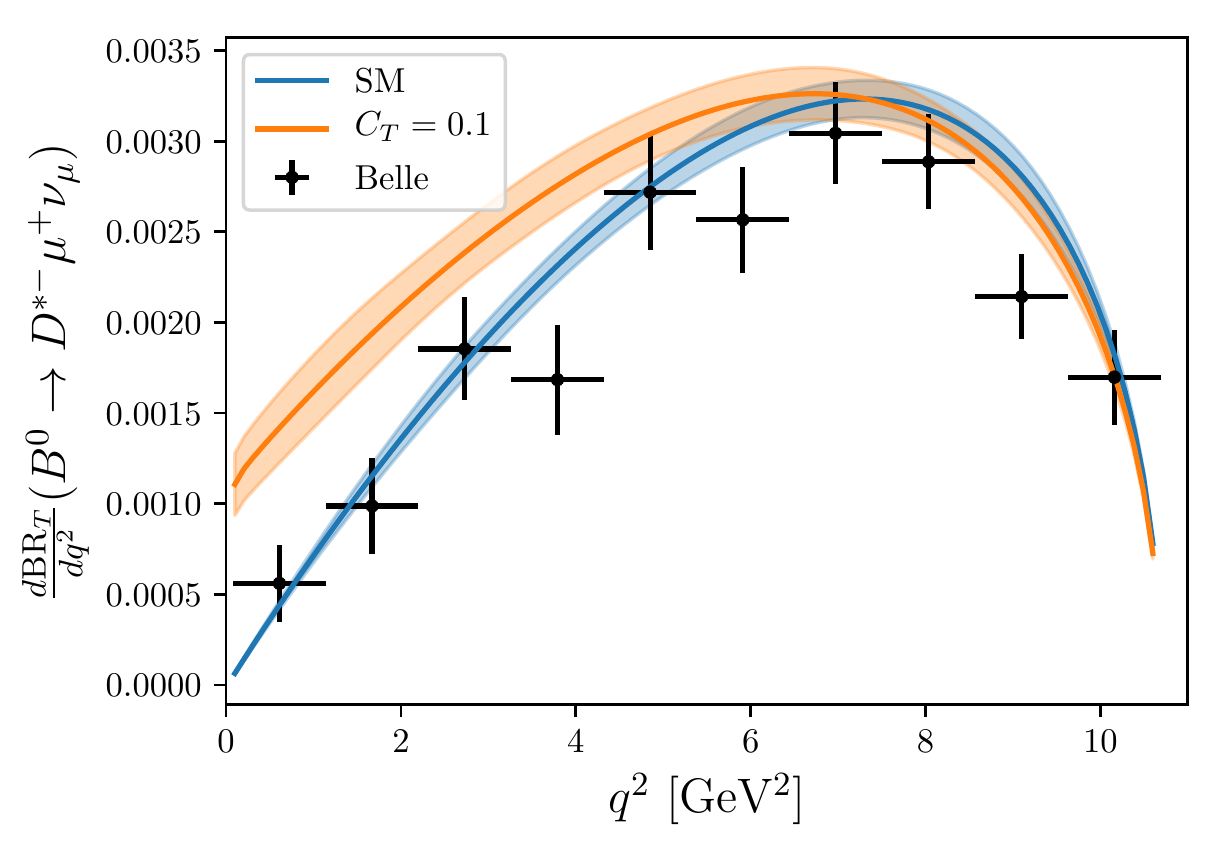}%
\includegraphics[width=0.45\textwidth]{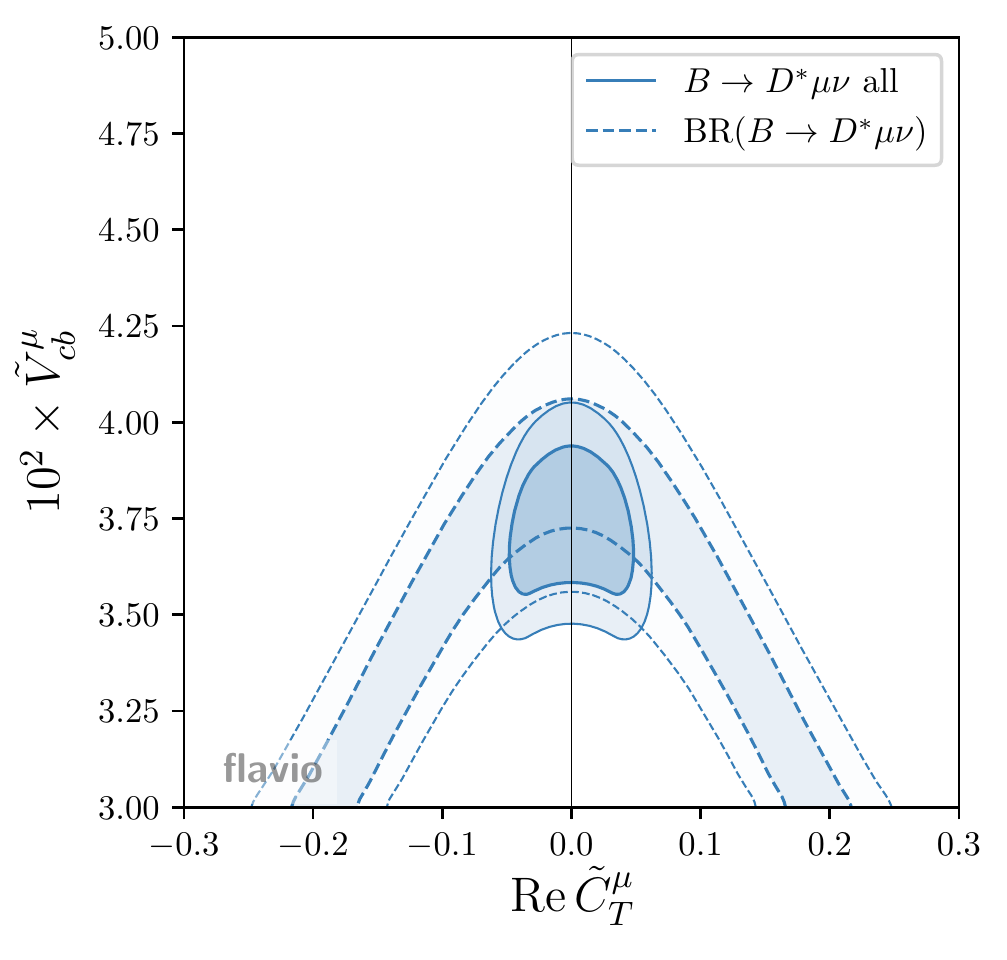}%
\caption{Left: Prediction for the transverse differential $B\to D^*\mu\nu$ branching ratio
in the SM (blue band) and a scenario with new physics in $C_T^\mu$ (orange
band) vs.\ the Belle measurement, demonstrating the different endpoint
behaviour at maximum recoil ($q^2=0$).
Both scenarios predict the same total $B\to D^*\mu\nu$ branching ratio.
Right: Comparison of the constraint on the tensor coefficient $\tilde C_T^\mu$ vs.\
$\tilde V_{cb}^{\mu}$ from the total $B\to D^*\mu\nu$ branching ratio measurements only (dashed) and using all $B\to D^*\mu\nu$
measurements (solid).}
\label{fig:CT_q2}
\end{figure}
As a consequence, this observable is exceptionally sensitive to tensor operators
near the maximum recoil point.
To illustrate this fact, in figure~\ref{fig:CT_q2} on the left we show the predictions
for the SM and a scenario
with a sizable NP contribution in $C_T$ compared to the experimental
data. The scenarios are chosen such that they give the same prediction
for the total branching ratio. The different behaviour at $q^2=0$
allows to clearly distinguish them and disfavours tensor contributions of this size. On the right of figure~\ref{fig:CT_q2} we
demonstrate again this qualitative difference by comparing the constraints from the total $B\to D^*\mu\nu$ rate alone and including
the differential distribution.

An additional observable that would be very sensitive to the tensor operator,
but has not been measured yet, is the ``flat term'' in $B\to D\ell\nu$. The
normalized differential decay rate as a function of the angle $\theta_\ell$ between the
charged lepton and the $B$ in the lepton-neutrino mass frame can be written
as
\begin{equation}
\frac{1}{\Gamma_\ell}\frac{d\Gamma_\ell}{d\cos\theta_\ell dq^2}
=
\frac{3}{4}\left[1-F_H(q^2)\right]\sin^2\theta_\ell
+\frac{1}{2}F_H(q^2)
+A_\text{FB}(q^2)\cos\theta_\ell \,.
\end{equation}
In complete analogy to the $B\to K\ell^+\ell^-$ decay, the observable $F_H$
vanishes in the SM up to tiny lepton mass effects, but can be sizable in the
presence of new physics in the tensor operator. Neglecting the lepton masses
and allowing for NP in $C_T$ and $C_{V_L}$, one finds
\begin{equation}
F_H(q^2)\approx \frac{18 q^2 f_T^2(q^2)}{m_B^2 f_+^2(q^2)}\frac{|C_T|^2}{|1+C_{V_L}|^2} \,.
\end{equation}

\begin{figure}[tbp]
\centering{
\includegraphics[width=0.5\textwidth]{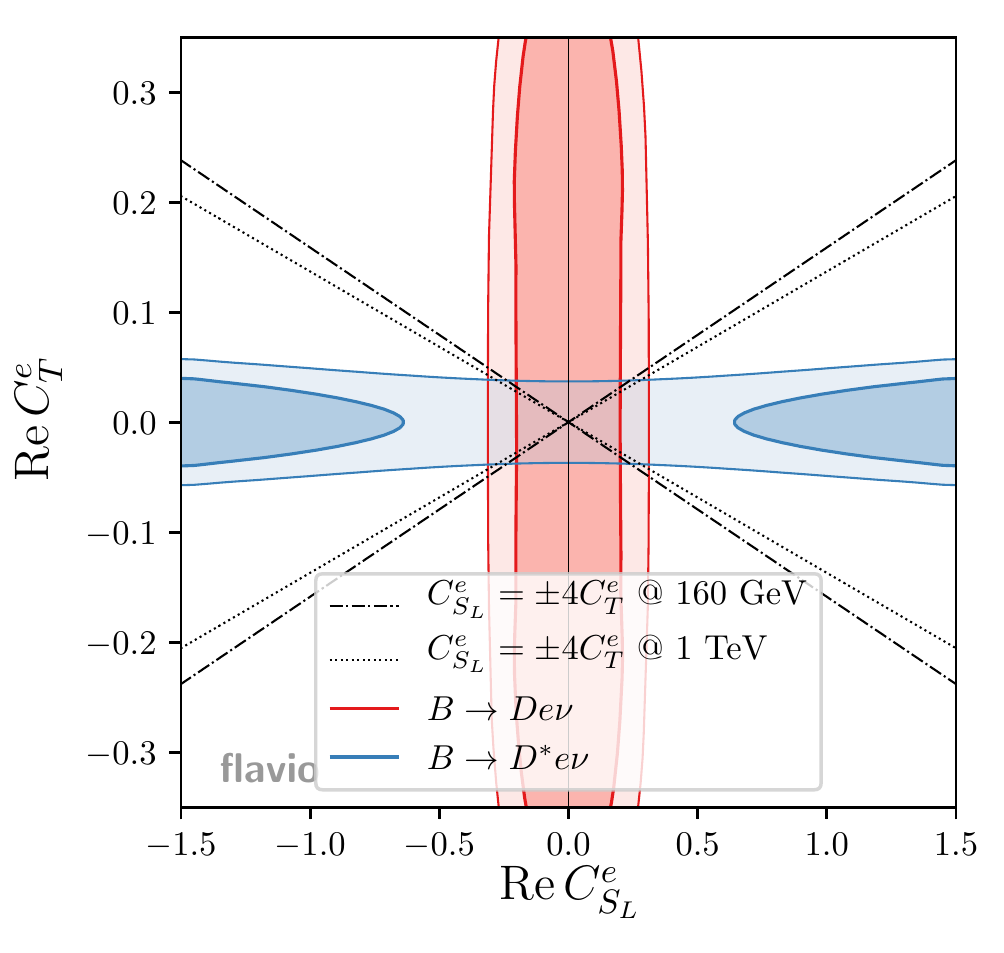}%
\includegraphics[width=0.5\textwidth]{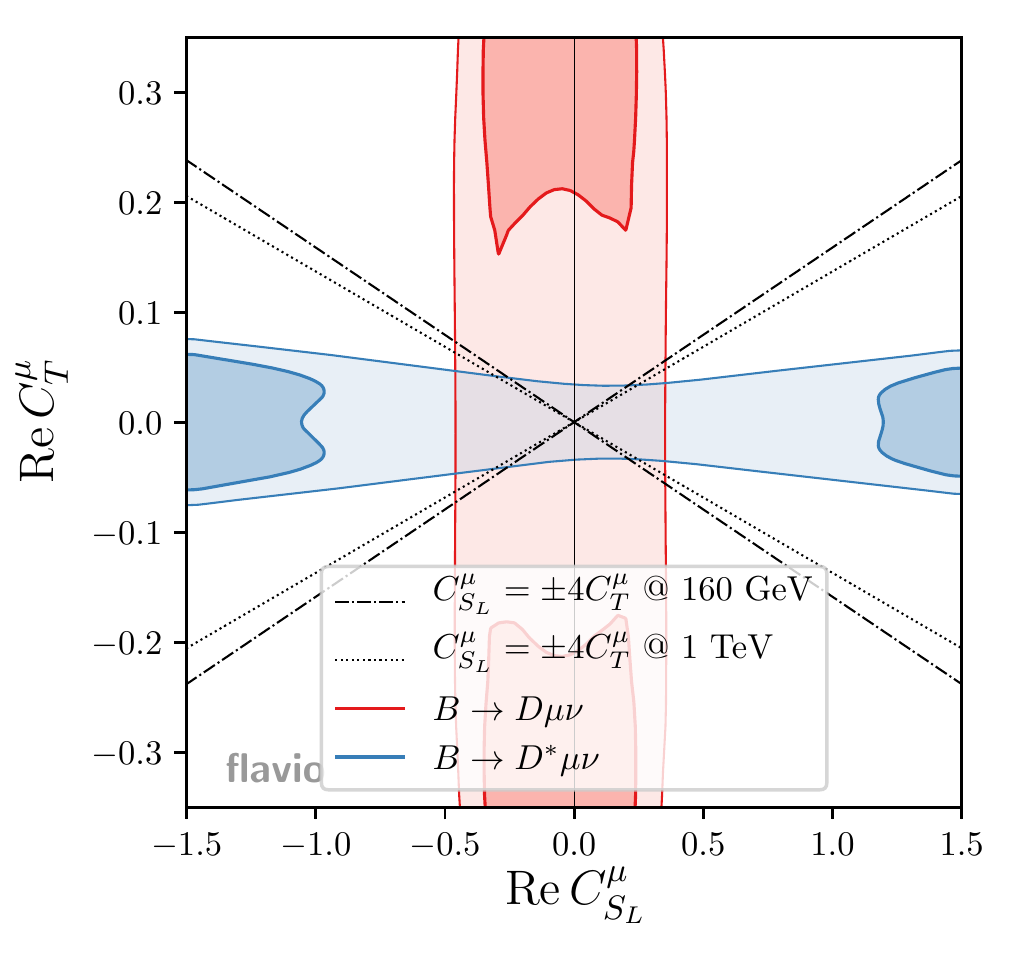}
}
\caption{Constraints on $C_{S_R}^{e,\mu}$ vs. $C_T^{e,\mu}$, profiling over $V_{cb}$.
The black dotted and dash-dotted lines show the expected ratio of the
Wilson coefficients in the two leptoquark scenarios
$S_1$ and $R_2$ (cf.\ table~\ref{tab:models}), assuming
the relation $C_{S_L}^\ell=\pm 4C_T^\ell$ to hold at the scale 160\,GeV
or 1\,TeV.}
\label{fig:CS_CT}
\end{figure}

Figure~\ref{fig:CS_CT}
shows the constraints on the tensor and left-handed scalar  operators, which always appear together in models with a tree-level
mediator, see Table~\ref{tab:models}, specifically in leptoquark models.  The displayed constraints from $B\to
D\ell\nu$ and $B\to D^*\ell\nu$, shown separately for electrons and muons, demonstrate clearly the strong sensitivity of $B\to
D^*\ell\nu$ to tensor contributions.
While the individual modes $B\to D^*e\nu$, $B\to D\mu\nu$, and $B\to D^*\mu\nu$
show a slight preference for non-zero NP contributions in either $C_{S_L}^\ell$
or $C_T^\ell$, the combination of $B\to D\ell\nu$ and $B\to D^*\ell\nu$ constraints
allows neither of these solutions and leads to a strong constraint on both operators.

\section{Conclusions}\label{sec:conc}

Semi-leptonic charged-current transitions $b\to c\ell\nu$ with $\ell=e$ or $\mu$
are traditionally used to measure the CKM element $V_{cb}$. In principle,
this transition could be affected by new physics with vector, scalar, or tensor
interactions, possibly violating lepton flavour universality.
This is motivated by the long-standing tensions between inclusive and exclusive determinations
of $V_{cb}$, but also by hints of a violation of lepton-flavour universality in $b\to c\tau\nu$ and $b\to s\ell\ell$
transitions.
We have conducted a comprehensive analysis of general new-physics effects in
$b\to c\ell\nu$ transitions, considering for the first time
the full operator basis and employing for the first time in a new physics analysis
measurements of $B\to D^*\ell\nu$ angular observables.

Our main findings can be summarized as follows.
\begin{itemize}
 \item Extracting the absolute value of the CKM element $V_{cb}$ from fits to the full sets of $B\to D\ell \nu$ and $B\to D^*\ell
 \nu$ data in Table~\ref{tab:data} yields values consistent with the recent literature
 and no significant tension between determinations from $B\to D\ell\nu$ vs.\ $B\to D^*\ell\nu$.
 \item We find no dependence of the $V_{cb}$ extraction on the statistical approach,
 but find a significant dependence on the treatment of systematic uncertainties in binned observables due to the ``d'Agostini bias''.
 \item We find that NP in right-handed currents cannot improve the agreement between inclusive and exclusive
 determinations of $V_{cb}$. Thanks to our use of differntial and angular dsitributions, this conclusion
 can even be drawn considering $B\to D^*\ell\nu$ vs.\ $B\to X_c\ell\nu$ alone.
 \item We find strong constraints on violations of $e$-$\mu$ universality, specifically for $C_{V_L}^\ell$.
 \item We demonstrate that the zero-recoil endpoint of the $B\to D\ell\nu$ spectrum is exceptionally sensitive to NP
 in scalar operators.
 \item We demonstrate that the maximum-recoil endpoint of the transverse $B\to D^*\ell\nu$ spectrum is exceptionally sensitive to NP
 in tensor operators.
\end{itemize}

Our analysis could be improved in several respects.
The treatment of experimental data had partly to rely on crude estimates
of the systematic uncertainties or correlations, where these were not public.
We urge the experimental collaborations to publish this information for
future and also existing analyses.
Our treatment of the inclusive decay is also approximate, as discussed at the
beginning of section~\ref{sec:np}. Clearly, a full fit to the moments of the
inclusive mode would be interesting, but is beyond the scope of our present
analysis.
Finally, our treatment of $B\to D^*$ form factors had to some extent to rely on the
heavy quark expansion, with only partial inclusion of $1/m_{c,b}^2$ contributions.
A full calculation of the $q^2$ dependence from lattice QCD, ideally including tensor form factors, would make these
constraints much more reliable.
We emphasize again that our analysis can be easily improved once this
information becomes available as all of our code is open source.

\section*{Acknowledgments}

We thank
Florian Bernlochner,
Christoph Bobeth,
Jorge Martin Camalich,
Danny van Dyk,
Paolo Gambino,
Thomas Kuhr,
Marcello Rotondo,
Stefan Schacht,
and
Christoph Schwanda
for useful discussions.
The work of M.~J.\  and D.~S.\ is supported by the DFG cluster of excellence ``Origin and
Structure of the Universe''.
The numerical analysis has been carried out on the computing facilities of the Computational Center for Particle and Astrophysics (C2PAP).

\appendix

\section{Numerical results for $B\to D^{(*)}$ form factors}

In this appendix, we give details on the parametrization of form factors discussed in section~\ref{sec:ff} and present the numerical results of our fit to lattice and LCSR calculations employing the HQET parametrization and unitarity bounds.
The $B\to D^{(*)}$ form factors $h_i$ in the HQET basis\footnote{The relation of the HQET form factors to the traditional form factor basis $V_i$, $A_i$, $T_i$ and $f_i$ can be found in appendix A of ref.~\cite{Sakaki:2013bfa}.
} can be found in ref.~\cite{Bernlochner:2017jka}. In the heavy quark limit, they vanish or reduce to a common form factor, the leading
Isgur-Wise function $\xi(w)$. It is thus convenient to write the form factors as $h_i(w) = \xi(w)\hat h(w)$. The expressions for all
$\hat h(w)$ at next-to-leading order in $\alpha_s$ and next-to-leading power in $\epsilon_{b,c}=\bar\Lambda/2m_{b,c}$ can be found in
ref.~\cite{Bernlochner:2017jka}. As discussed in section~\ref{sec:ff}, we modify these expressions by allowing for an additional
$O(\epsilon_c^2)$ correction to the form factors that are protected from $O(\epsilon_c)$ corrections:
\begin{align}
\hat h_{A_1}(w) &\to \hat h_{A_1}(w) + \epsilon_c^2\delta_{h_{A_1}}
\,,&
\hat h_{T_1}(w) &\to \hat h_{T_1}(w) + \epsilon_c^2\delta_{h_{T_1}}
\,,&
\hat h_{+}(w) &\to \hat h_{+}(w) + \epsilon_c^2\delta_{h_{+}}
\,.&
\end{align}
We neglect a possible $w$ dependence of the $\delta_i$ terms.
The leading order Isgur-Wise function $\xi$ can be written to second order in the $z$ expansion as
\begin{equation}
\xi(z) = 1 - 8  \rho^2  z + (64  c - 16  \rho^2)  z^2.
\end{equation}
We then perform a Bayesian fit (a Markov Chain Monte Carlo employing flavio \cite{Straub:2018kue} and emcee \cite{ForemanMackey:2012ig}) to the
theory constraints described in section~\ref{sec:ff} of the ten parameters parametrizing the functions $\hat h_i$ and
$\xi$.\footnote{Again these values have also been obtained by an independent frequentist implementation.} The mean, standard deviation
and correlation matrix of nine of those parameters is
\begin{gather}
\begin{pmatrix}
\chi_2(1) \\
\chi_2'(1) \\
\chi_3'(1) \\
\eta(1) \\
\eta'(1) \\
\rho^2 \\
c \\
\delta_{h_{A_1}} \\
\delta_{h_+}
\end{pmatrix}
=
\begin{pmatrix}
-0.058 \pm 0.019 \\
-0.001 \pm 0.020 \\
0.035 \pm 0.019 \\
0.358 \pm 0.043 \\
0.044 \pm 0.125 \\
1.306 \pm 0.059 \\
1.220 \pm 0.109 \\
-2.299 \pm 0.394 \\
0.485 \pm 0.269
\end{pmatrix}
\\
\rho =
\begin{pmatrix}
  1.00 & 0.01 & 0.02 & -0.00 & 0.02 & -0.27 & -0.21 & -0.03 & 0.02 \\
   0.01 & 1.00 & -0.00 & -0.02 & -0.02 & 0.00 & 0.14 & 0.01 & 0.00 \\
   0.02 & -0.00 & 1.00 & 0.00 & -0.03 & 0.83 & 0.61 & -0.03 & 0.02 \\
   -0.00 & -0.02 & 0.00 & 1.00 & 0.03 & 0.01 & 0.04 & 0.15 & 0.21 \\
   0.02 & -0.02 & -0.03 & 0.03 & 1.00 & -0.14 & -0.16 & -0.05 & -0.22 \\
   -0.27 & 0.00 & 0.83 & 0.01 & -0.14 & 1.00 & 0.79 & 0.09 & -0.14 \\
   -0.21 & 0.14 & 0.61 & 0.04 & -0.16 & 0.79 & 1.00 & 0.06 & -0.08 \\
   -0.03 & 0.01 & -0.03 & 0.15 & -0.05 & 0.09 & 0.06 & 1.00 & -0.24 \\
   0.02 & 0.00 & 0.02 & 0.21 & -0.22 & -0.14 & -0.08 & -0.24 & 1.00
\end{pmatrix}.
\end{gather}
The tenth parameter, $\delta_{T_1}$, is not constrained by the fit, thus its posterior is equal to the prior, which we conservatively take to be a Gaussian with mean 3 centered around 0.
These form factors have been implemented and set as defaults in flavio version 0.26.

\bibliographystyle{JHEP}
\bibliography{bibliography}

\providecommand{\href}[2]{#2}\begingroup\raggedright\begin{thebibliography}{10}

\bibitem{Charles:2015gya}
J.~Charles et~al., \textit{{Current status of the Standard Model CKM fit and
  constraints on $\Delta F=2$ New Physics}},  {\em Phys. Rev.} \textbf{D91}
  (2015), no.~7 073007,
  [\href{https://arxiv.org/abs/1501.05013}{\texttt{arXiv:1501.05013}}].

\bibitem{Bona:2017cxr}
\textbf{UTfit} Collaboration, M.~Bona, \textit{{Latest results for the Unitary
  Triangle fit from the UTfit Collaboration}},  {\em PoS} \textbf{CKM2016}
  (2017) 096.

\bibitem{Alberti:2014yda}
A.~Alberti, P.~Gambino, K.~J. Healey, and S.~Nandi, \textit{{Precision
  Determination of the Cabibbo-Kobayashi-Maskawa Element $V_{cb}$}},  {\em
  Phys. Rev. Lett.} \textbf{114} (2015), no.~6 061802,
  [\href{https://arxiv.org/abs/1411.6560}{\texttt{arXiv:1411.6560}}].

\bibitem{Ricciardi:2017lne}
G.~Ricciardi, \textit{{Semileptonic decays and $|V_{xb}|$ determinations}},
  2017.
\newblock \href{https://arxiv.org/abs/1712.06988}{\texttt{arXiv:1712.06988}}.

\bibitem{Na:2015kha}
\textbf{HPQCD} Collaboration, H.~Na, C.~M. Bouchard, G.~P. Lepage, C.~Monahan,
  and J.~Shigemitsu, \textit{{$B \rightarrow D l \nu$ form factors at nonzero
  recoil and extraction of $|V_{cb}|$}},  {\em Phys. Rev.} \textbf{D92} (2015),
  no.~5 054510,
  [\href{https://arxiv.org/abs/1505.03925}{\texttt{arXiv:1505.03925}}].
  [Erratum: Phys. Rev.D93,no.11,119906(2016)].

\bibitem{Lattice:2015rga}
\textbf{MILC} Collaboration, J.~A. Bailey et~al., \textit{{$B\to D\ell\nu$ form
  factors at nonzero recoil and $|V_{cb}|$ from 2+1-flavor lattice QCD}},  {\em
  Phys. Rev.} \textbf{D92} (2015), no.~3 034506,
  [\href{https://arxiv.org/abs/1503.07237}{\texttt{arXiv:1503.07237}}].

\bibitem{Bailey:2014tva}
\textbf{Fermilab Lattice, MILC} Collaboration, J.~A. Bailey et~al.,
  \textit{{Update of $|V_{cb}|$ from the $\bar{B}\to D^*\ell\bar{\nu}$ form
  factor at zero recoil with three-flavor lattice QCD}},  {\em Phys. Rev.}
  \textbf{D89} (2014), no.~11 114504,
  [\href{https://arxiv.org/abs/1403.0635}{\texttt{arXiv:1403.0635}}].

\bibitem{Harrison:2017fmw}
J.~Harrison, C.~Davies, and M.~Wingate, \textit{{Lattice QCD calculation of the
  ${{B}_{(s)}\to D_{(s)}^{*}\ell{\nu}}$ form factors at zero recoil and
  implications for ${|V_{cb}|}$}},
  \href{https://arxiv.org/abs/1711.11013}{\texttt{arXiv:1711.11013}}.

\bibitem{Bigi:2016mdz}
D.~Bigi and P.~Gambino, \textit{{Revisiting $B\to D \ell \nu$}},  {\em Phys.
  Rev.} \textbf{D94} (2016), no.~9 094008,
  [\href{https://arxiv.org/abs/1606.08030}{\texttt{arXiv:1606.08030}}].

\bibitem{Bigi:2017njr}
D.~Bigi, P.~Gambino, and S.~Schacht, \textit{{A fresh look at the determination
  of $|V_{cb}|$ from $B\to D^{*} \ell \nu$}},  {\em Phys. Lett.} \textbf{B769}
  (2017) 441--445,
  [\href{https://arxiv.org/abs/1703.06124}{\texttt{arXiv:1703.06124}}].

\bibitem{Bigi:2017jbd}
D.~Bigi, P.~Gambino, and S.~Schacht, \textit{{$R(D^*)$, $|V_{cb}|$, and the
  Heavy Quark Symmetry relations between form factors}},  {\em JHEP}
  \textbf{11} (2017) 061,
  [\href{https://arxiv.org/abs/1707.09509}{\texttt{arXiv:1707.09509}}].

\bibitem{Aoki:2016frl}
S.~Aoki et~al., \textit{{Review of lattice results concerning low-energy
  particle physics}},  {\em Eur. Phys. J.} \textbf{C77} (2017), no.~2 112,
  [\href{https://arxiv.org/abs/1607.00299}{\texttt{arXiv:1607.00299}}].

\bibitem{Bernlochner:2017jka}
F.~U. Bernlochner, Z.~Ligeti, M.~Papucci, and D.~J. Robinson, \textit{{Combined
  analysis of semileptonic $B$ decays to $D$ and $D^*$: $R(D^{(*)})$,
  $|V_{cb}|$, and new physics}},  {\em Phys. Rev.} \textbf{D95} (2017), no.~11
  115008, [\href{https://arxiv.org/abs/1703.05330}{\texttt{arXiv:1703.05330}}].

\bibitem{Grinstein:2017nlq}
B.~Grinstein and A.~Kobach, \textit{{Model-Independent Extraction of $|V_{cb}|$
  from $\bar{B}\rightarrow D^* \ell \overline{\nu}$}},  {\em Phys. Lett.}
  \textbf{B771} (2017) 359--364,
  [\href{https://arxiv.org/abs/1703.08170}{\texttt{arXiv:1703.08170}}].

\bibitem{Bernlochner:2017xyx}
F.~U. Bernlochner, Z.~Ligeti, M.~Papucci, and D.~J. Robinson, \textit{{Tensions
  and correlations in $|V_{cb}|$ determinations}},  {\em Phys. Rev.}
  \textbf{D96} (2017), no.~9 091503,
  [\href{https://arxiv.org/abs/1708.07134}{\texttt{arXiv:1708.07134}}].

\bibitem{Jaiswal:2017rve}
S.~Jaiswal, S.~Nandi, and S.~K. Patra, \textit{{Extraction of $|V_{cb}|$ from
  $B\to D^{(*)}\ell\nu_\ell$ and the Standard Model predictions of
  $R(D^{(*)})$}},  {\em JHEP} \textbf{12} (2017) 060,
  [\href{https://arxiv.org/abs/1707.09977}{\texttt{arXiv:1707.09977}}].

\bibitem{Lees:2013uzd}
\textbf{BaBar} Collaboration, J.~P. Lees et~al., \textit{{Measurement of an
  Excess of $\bar{B} \to D^{(*)}\tau^- \bar{\nu}_\tau$ Decays and Implications
  for Charged Higgs Bosons}},  {\em Phys. Rev.} \textbf{D88} (2013), no.~7
  072012, [\href{https://arxiv.org/abs/1303.0571}{\texttt{arXiv:1303.0571}}].

\bibitem{Huschle:2015rga}
\textbf{Belle} Collaboration, M.~Huschle et~al., \textit{{Measurement of the
  branching ratio of $\bar{B} \to D^{(\ast)} \tau^- \bar{\nu}_\tau$ relative to
  $\bar{B} \to D^{(\ast)} \ell^- \bar{\nu}_\ell$ decays with hadronic tagging
  at Belle}},  {\em Phys. Rev.} \textbf{D92} (2015), no.~7 072014,
  [\href{https://arxiv.org/abs/1507.03233}{\texttt{arXiv:1507.03233}}].

\bibitem{Aaij:2015yra}
\textbf{LHCb} Collaboration, R.~Aaij et~al., \textit{{Measurement of the ratio
  of branching fractions $\mathcal{B}(\bar{B}^0 \to
  D^{*+}\tau^{-}\bar{\nu}_{\tau})/\mathcal{B}(\bar{B}^0 \to
  D^{*+}\mu^{-}\bar{\nu}_{\mu})$}},  {\em Phys. Rev. Lett.} \textbf{115}
  (2015), no.~11 111803,
  [\href{https://arxiv.org/abs/1506.08614}{\texttt{arXiv:1506.08614}}].
  [Erratum: Phys. Rev. Lett.115,no.15,159901(2015)].

\bibitem{Sato:2016svk}
\textbf{Belle} Collaboration, Y.~Sato et~al., \textit{{Measurement of the
  branching ratio of $\bar{B}^0 \rightarrow D^{*+} \tau^- \bar{\nu}_{\tau}$
  relative to $\bar{B}^0 \rightarrow D^{*+} \ell^- \bar{\nu}_{\ell}$ decays
  with a semileptonic tagging method}},  {\em Phys. Rev.} \textbf{D94} (2016),
  no.~7 072007,
  [\href{https://arxiv.org/abs/1607.07923}{\texttt{arXiv:1607.07923}}].

\bibitem{Hirose:2016wfn}
\textbf{Belle} Collaboration, S.~Hirose et~al., \textit{{Measurement of the
  $\tau$ lepton polarization and $R(D^*)$ in the decay $\bar{B} \to D^* \tau^-
  \bar{\nu}_\tau$}},  {\em Phys. Rev. Lett.} \textbf{118} (2017), no.~21
  211801, [\href{https://arxiv.org/abs/1612.00529}{\texttt{arXiv:1612.00529}}].

\bibitem{Aaij:2017uff}
\textbf{LHCb} Collaboration, R.~Aaij et~al., \textit{{Measurement of the ratio
  of the $B^0 \to D^{*-} \tau^+ \nu_{\tau}$ and $B^0 \to D^{*-} \mu^+
  \nu_{\mu}$ branching fractions using three-prong $\tau$-lepton decays}},
  \href{https://arxiv.org/abs/1708.08856}{\texttt{arXiv:1708.08856}}.

\bibitem{Aaij:2017tyk}
\textbf{LHCb} Collaboration, R.~Aaij et~al., \textit{{Measurement of the ratio
  of branching fractions
  $\mathcal{B}(B_c^+\,\to\,J/\psi\tau^+\nu_\tau)$/$\mathcal{B}(B_c^+\,\to\,J/\psi\mu^+\nu_\mu)$}},
  \href{https://arxiv.org/abs/1711.05623}{\texttt{arXiv:1711.05623}}.

\bibitem{Bauer:2015knc}
M.~Bauer and M.~Neubert, \textit{{Minimal Leptoquark Explanation for the
  R$_{D^{(*)}}$ , R$_K$ , and $(g-2)_\mu$ Anomalies}},  {\em Phys. Rev. Lett.}
  \textbf{116} (2016), no.~14 141802,
  [\href{https://arxiv.org/abs/1511.01900}{\texttt{arXiv:1511.01900}}].

\bibitem{Becirevic:2016oho}
D.~Bečirević, N.~Košnik, O.~Sumensari, and R.~Zukanovich~Funchal,
  \textit{{Palatable Leptoquark Scenarios for Lepton Flavor Violation in
  Exclusive $b\to s\ell_1\ell_2$ modes}},  {\em JHEP} \textbf{11} (2016) 035,
  [\href{https://arxiv.org/abs/1608.07583}{\texttt{arXiv:1608.07583}}].

\bibitem{Cai:2017wry}
Y.~Cai, J.~Gargalionis, M.~A. Schmidt, and R.~R. Volkas, \textit{{Reconsidering
  the One Leptoquark solution: flavor anomalies and neutrino mass}},  {\em
  JHEP} \textbf{10} (2017) 047,
  [\href{https://arxiv.org/abs/1704.05849}{\texttt{arXiv:1704.05849}}].

\bibitem{Buttazzo:2017ixm}
D.~Buttazzo, A.~Greljo, G.~Isidori, and D.~Marzocca, \textit{{B-physics
  anomalies: a guide to combined explanations}},  {\em JHEP} \textbf{11} (2017)
  044, [\href{https://arxiv.org/abs/1706.07808}{\texttt{arXiv:1706.07808}}].

\bibitem{Dassinger:2007pj}
B.~M. Dassinger, R.~Feger, and T.~Mannel, \textit{{Testing the left-handedness
  of the $b\to c$ transition}},  {\em Phys. Rev.} \textbf{D75} (2007) 095007,
  [\href{https://arxiv.org/abs/hep-ph/0701054}{\texttt{hep-ph/0701054}}].

\bibitem{Dassinger:2008as}
B.~Dassinger, R.~Feger, and T.~Mannel, \textit{{Complete Michel Parameter
  Analysis of inclusive semileptonic $b \to c$ transition}},  {\em Phys. Rev.}
  \textbf{D79} (2009) 075015,
  [\href{https://arxiv.org/abs/0803.3561}{\texttt{arXiv:0803.3561}}].

\bibitem{Crivellin:2009sd}
A.~Crivellin, \textit{{Effects of right-handed charged currents on the
  determinations of $|V_{ub}|$ and $|V_{cb}|$}},  {\em Phys. Rev.} \textbf{D81}
  (2010) 031301,
  [\href{https://arxiv.org/abs/0907.2461}{\texttt{arXiv:0907.2461}}].

\bibitem{Feger:2010qc}
R.~Feger, T.~Mannel, V.~Klose, H.~Lacker, and T.~Luck, \textit{{Limit on a
  Right-Handed Admixture to the Weak $b \to c$ Current from Semileptonic
  Decays}},  {\em Phys. Rev.} \textbf{D82} (2010) 073002,
  [\href{https://arxiv.org/abs/1003.4022}{\texttt{arXiv:1003.4022}}].

\bibitem{Faller:2011nj}
S.~Faller, T.~Mannel, and S.~Turczyk, \textit{{Limits on New Physics from
  exclusive $B \to D^{(*)}\ell \bar\nu$ Decays}},  {\em Phys. Rev.}
  \textbf{D84} (2011) 014022,
  [\href{https://arxiv.org/abs/1105.3679}{\texttt{arXiv:1105.3679}}].

\bibitem{Crivellin:2014zpa}
A.~Crivellin and S.~Pokorski, \textit{{Can the differences in the
  determinations of $V_{ub}$ and $V_{cb}$ be explained by New Physics?}},  {\em
  Phys. Rev. Lett.} \textbf{114} (2015), no.~1 011802,
  [\href{https://arxiv.org/abs/1407.1320}{\texttt{arXiv:1407.1320}}].

\bibitem{Colangelo:2016ymy}
P.~Colangelo and F.~De~Fazio, \textit{{Tension in the inclusive versus
  exclusive determinations of $|V_{cb}|$: a possible role of new physics}},
  {\em Phys. Rev.} \textbf{D95} (2017), no.~1 011701,
  [\href{https://arxiv.org/abs/1611.07387}{\texttt{arXiv:1611.07387}}].

\bibitem{Aubert:2009ac}
\textbf{BaBar} Collaboration, B.~Aubert et~al., \textit{{Measurement of
  $|V_{cb}|$ and the Form-Factor Slope in $\bar B \to D l^- \bar\nu$ Decays in
  Events Tagged by a Fully Reconstructed B Meson}},  {\em Phys. Rev. Lett.}
  \textbf{104} (2010) 011802,
  [\href{https://arxiv.org/abs/0904.4063}{\texttt{arXiv:0904.4063}}].

\bibitem{Dungel:2010uk}
\textbf{Belle} Collaboration, W.~Dungel et~al., \textit{{Measurement of the
  form factors of the decay $B^0 \to D^* \ell^+ \nu$ and determination of the
  CKM matrix element $|V_{cb}|$}},  {\em Phys. Rev.} \textbf{D82} (2010)
  112007, [\href{https://arxiv.org/abs/1010.5620}{\texttt{arXiv:1010.5620}}].

\bibitem{Glattauer:2015teq}
\textbf{Belle} Collaboration, R.~Glattauer et~al., \textit{{Measurement of the
  decay $B\to D\ell\nu_\ell$ in fully reconstructed events and determination of
  the Cabibbo-Kobayashi-Maskawa matrix element $|V_{cb}|$}},  {\em Phys. Rev.}
  \textbf{D93} (2016), no.~3 032006,
  [\href{https://arxiv.org/abs/1510.03657}{\texttt{arXiv:1510.03657}}].

\bibitem{Abdesselam:2017kjf}
\textbf{Belle} Collaboration, A.~Abdesselam et~al., \textit{{Precise
  determination of the CKM matrix element $\left| V_{cb}\right|$ with $\bar B^0
  \to D^{*\,+} \, \ell^- \, \bar \nu_\ell$ decays with hadronic tagging at
  Belle}},  \href{https://arxiv.org/abs/1702.01521}{\texttt{arXiv:1702.01521}}.

\bibitem{Straub:2018kue}
D.~M. Straub, \textit{{flavio: a Python package for flavour and precision
  phenomenology in the Standard Model and beyond}},
  \href{https://arxiv.org/abs/1810.08132}{\texttt{arXiv:1810.08132}}.
  \url{https://flav-io.github.io}.

\bibitem{Goldberger:1999yh}
W.~D. Goldberger, \textit{{Semileptonic B decays as a probe of new physics}},
  \href{https://arxiv.org/abs/hep-ph/9902311}{\texttt{hep-ph/9902311}}.

\bibitem{Buchmuller:1985jz}
W.~Buchm{\"u}ller and D.~Wyler, \textit{{Effective Lagrangian Analysis of New
  Interactions and Flavor Conservation}},  {\em Nucl.~Phys.} \textbf{B268}
  (1986) 621--653.

\bibitem{Grzadkowski:2010es}
B.~Grzadkowski, M.~Iskrzynski, M.~Misiak, and J.~Rosiek, \textit{{Dimension-Six
  Terms in the Standard Model Lagrangian}},  {\em JHEP} \textbf{1010} (2010)
  085, [\href{https://arxiv.org/abs/1008.4884}{\texttt{arXiv:1008.4884}}].

\bibitem{Cirigliano:2012ab}
V.~Cirigliano, M.~Gonzalez-Alonso, and M.~L. Graesser, \textit{{Non-standard
  Charged Current Interactions: beta decays versus the LHC}},  {\em JHEP}
  \textbf{02} (2013) 046,
  [\href{https://arxiv.org/abs/1210.4553}{\texttt{arXiv:1210.4553}}].

\bibitem{Alonso:2015sja}
R.~Alonso, B.~Grinstein, and J.~Martin~Camalich, \textit{{Lepton universality
  violation and lepton flavor conservation in $B$-meson decays}},  {\em JHEP}
  \textbf{10} (2015) 184,
  [\href{https://arxiv.org/abs/1505.05164}{\texttt{arXiv:1505.05164}}].

\bibitem{Cata:2015lta}
O.~Catà and M.~Jung, \textit{{Signatures of a nonstandard Higgs boson from
  flavor physics}},  {\em Phys. Rev.} \textbf{D92} (2015), no.~5 055018,
  [\href{https://arxiv.org/abs/1505.05804}{\texttt{arXiv:1505.05804}}].

\bibitem{Aebischer:2015fzz}
J.~Aebischer, A.~Crivellin, M.~Fael, and C.~Greub, \textit{{Matching of gauge
  invariant dimension-six operators for $b\to s$ and $b\to c$ transitions}},
  {\em JHEP} \textbf{05} (2016) 037,
  [\href{https://arxiv.org/abs/1512.02830}{\texttt{arXiv:1512.02830}}].

\bibitem{Cirigliano:2009wk}
V.~Cirigliano, J.~Jenkins, and M.~Gonzalez-Alonso, \textit{{Semileptonic decays
  of light quarks beyond the Standard Model}},  {\em Nucl. Phys.} \textbf{B830}
  (2010) 95--115,
  [\href{https://arxiv.org/abs/0908.1754}{\texttt{arXiv:0908.1754}}].

\bibitem{Aloni:2017eny}
D.~Aloni, A.~Efrati, Y.~Grossman, and Y.~Nir, \textit{{$\Upsilon$ and $\psi$
  leptonic decays as probes of solutions to the $R_D^{(*)}$ puzzle}},  {\em
  JHEP} \textbf{06} (2017) 019,
  [\href{https://arxiv.org/abs/1702.07356}{\texttt{arXiv:1702.07356}}].

\bibitem{Jenkins:2013zja}
E.~E. Jenkins, A.~V. Manohar, and M.~Trott, \textit{{Renormalization Group
  Evolution of the Standard Model Dimension Six Operators I: Formalism and
  lambda Dependence}},  {\em JHEP} \textbf{10} (2013) 087,
  [\href{https://arxiv.org/abs/1308.2627}{\texttt{arXiv:1308.2627}}].

\bibitem{Jenkins:2013wua}
E.~E. Jenkins, A.~V. Manohar, and M.~Trott, \textit{{Renormalization Group
  Evolution of the Standard Model Dimension Six Operators II: Yukawa
  Dependence}},  {\em JHEP} \textbf{01} (2014) 035,
  [\href{https://arxiv.org/abs/1310.4838}{\texttt{arXiv:1310.4838}}].

\bibitem{Alonso:2013hga}
R.~Alonso, E.~E. Jenkins, A.~V. Manohar, and M.~Trott, \textit{{Renormalization
  Group Evolution of the Standard Model Dimension Six Operators III: Gauge
  Coupling Dependence and Phenomenology}},  {\em JHEP} \textbf{04} (2014) 159,
  [\href{https://arxiv.org/abs/1312.2014}{\texttt{arXiv:1312.2014}}].

\bibitem{Faller:2008tr}
S.~Faller, A.~Khodjamirian, C.~Klein, and T.~Mannel, \textit{{$B \to D^{(*)}$
  Form Factors from QCD Light-Cone Sum Rules}},  {\em Eur. Phys. J.}
  \textbf{C60} (2009) 603--615,
  [\href{https://arxiv.org/abs/0809.0222}{\texttt{arXiv:0809.0222}}].

\bibitem{Shifman:1986sm}
M.~A. Shifman and M.~B. Voloshin, \textit{{On Annihilation of Mesons Built from
  Heavy and Light Quark and $\bar B^0 \leftrightarrow B^0$ Oscillations}},
  {\em Sov. J. Nucl. Phys.} \textbf{45} (1987) 292. [Yad. Fiz.45,463(1987)].

\bibitem{Isgur:1989vq}
N.~Isgur and M.~B. Wise, \textit{{Weak Decays of Heavy Mesons in the Static
  Quark Approximation}},  {\em Phys. Lett.} \textbf{B232} (1989) 113--117.

\bibitem{Isgur:1989ed}
N.~Isgur and M.~B. Wise, \textit{{WEAK TRANSITION FORM-FACTORS BETWEEN HEAVY
  MESONS}},  {\em Phys. Lett.} \textbf{B237} (1990) 527--530.

\bibitem{Luke:1990eg}
M.~E. Luke, \textit{{Effects of subleading operators in the heavy quark
  effective theory}},  {\em Phys. Lett.} \textbf{B252} (1990) 447--455.

\bibitem{Neubert:1991xw}
M.~Neubert and V.~Rieckert, \textit{{New approach to the universal form-factors
  in decays of heavy mesons}},  {\em Nucl. Phys.} \textbf{B382} (1992) 97--119.

\bibitem{Falk:1992wt}
A.~F. Falk and M.~Neubert, \textit{{Second order power corrections in the heavy
  quark effective theory. 1. Formalism and meson form-factors}},  {\em Phys.
  Rev.} \textbf{D47} (1993) 2965--2981,
  [\href{https://arxiv.org/abs/hep-ph/9209268}{\texttt{hep-ph/9209268}}].

\bibitem{Neubert:1992qq}
M.~Neubert, \textit{{Renormalization of heavy quark currents}},  {\em Nucl.
  Phys.} \textbf{B371} (1992) 149--176.

\bibitem{Neubert:1992wq}
M.~Neubert, Z.~Ligeti, and Y.~Nir, \textit{{QCD sum rule analysis of the
  subleading Isgur-Wise form-factor $\chi_2 (v\cdot v')$}},  {\em Phys. Lett.}
  \textbf{B301} (1993) 101--107,
  [\href{https://arxiv.org/abs/hep-ph/9209271}{\texttt{hep-ph/9209271}}].

\bibitem{Neubert:1992pn}
M.~Neubert, Z.~Ligeti, and Y.~Nir, \textit{{The Subleading Isgur-Wise
  form-factor $\chi_3(v \cdot v'$) to order $\alpha_s$ in QCD sum rules}},
  {\em Phys. Rev.} \textbf{D47} (1993) 5060--5066,
  [\href{https://arxiv.org/abs/hep-ph/9212266}{\texttt{hep-ph/9212266}}].

\bibitem{Ligeti:1993hw}
Z.~Ligeti, Y.~Nir, and M.~Neubert, \textit{{The Subleading Isgur-Wise
  form-factor $\xi_3(v - v')$ and its implications for the decays $\bar B \to
  D^* \ell\bar\nu$}},  {\em Phys. Rev.} \textbf{D49} (1994) 1302--1309,
  [\href{https://arxiv.org/abs/hep-ph/9305304}{\texttt{hep-ph/9305304}}].

\bibitem{Czarnecki:1996gu}
A.~Czarnecki, \textit{{Two loop QCD corrections to $b \to c$ transitions at
  zero recoil}},  {\em Phys. Rev. Lett.} \textbf{76} (1996) 4124--4127,
  [\href{https://arxiv.org/abs/hep-ph/9603261}{\texttt{hep-ph/9603261}}].

\bibitem{Czarnecki:1997cf}
A.~Czarnecki and K.~Melnikov, \textit{{Two loop QCD corrections to $b \to c$
  transitions at zero recoil: Analytical results}},  {\em Nucl. Phys.}
  \textbf{B505} (1997) 65--83,
  [\href{https://arxiv.org/abs/hep-ph/9703277}{\texttt{hep-ph/9703277}}].

\bibitem{Boyd:1997kz}
C.~G. Boyd, B.~Grinstein, and R.~F. Lebed, \textit{{Precision corrections to
  dispersive bounds on form-factors}},  {\em Phys. Rev.} \textbf{D56} (1997)
  6895--6911,
  [\href{https://arxiv.org/abs/hep-ph/9705252}{\texttt{hep-ph/9705252}}].

\bibitem{Caprini:1997mu}
I.~Caprini, L.~Lellouch, and M.~Neubert, \textit{{Dispersive bounds on the
  shape of $\bar B \to D^{(*)}\ell\bar\nu$ form-factors}},  {\em Nucl. Phys.}
  \textbf{B530} (1998) 153--181,
  [\href{https://arxiv.org/abs/hep-ph/9712417}{\texttt{hep-ph/9712417}}].

\bibitem{Aubert:2008yv}
\textbf{BaBar} Collaboration, B.~Aubert et~al., \textit{{Measurements of the
  Semileptonic Decays $\bar B \to D l \bar\nu$ and $\bar B \to D* l \bar\nu$
  Using a Global Fit to $D X l \bar\nu$ Final States}},  {\em Phys. Rev.}
  \textbf{D79} (2009) 012002,
  [\href{https://arxiv.org/abs/0809.0828}{\texttt{arXiv:0809.0828}}].

\bibitem{Aubert:2007qw}
\textbf{BaBar} Collaboration, B.~Aubert et~al., \textit{{A Measurement of the
  branching fractions of exclusive $\bar{B} \to D^{(*)}$ ($\pi$) $\ell^{-}
  \bar{\nu}_\ell$ decays in events with a fully reconstructed $B$ meson}},
  {\em Phys. Rev. Lett.} \textbf{100} (2008) 151802,
  [\href{https://arxiv.org/abs/0712.3503}{\texttt{arXiv:0712.3503}}].

\bibitem{Aubert:2007rs}
\textbf{BaBar} Collaboration, B.~Aubert et~al., \textit{{Determination of the
  form-factors for the decay $B^0 \to D^{*-} \ell^{+} \nu_{l}$ and of the CKM
  matrix element $|V_{cb}|$}},  {\em Phys. Rev.} \textbf{D77} (2008) 032002,
  [\href{https://arxiv.org/abs/0705.4008}{\texttt{arXiv:0705.4008}}].

\bibitem{Aubert:2007qs}
\textbf{BaBar} Collaboration, B.~Aubert et~al., \textit{{Measurement of the
  Decay $B^{-} \to$ D*0 $e^{-} \bar{\nu}$( $e$)}},  {\em Phys. Rev. Lett.}
  \textbf{100} (2008) 231803,
  [\href{https://arxiv.org/abs/0712.3493}{\texttt{arXiv:0712.3493}}].

\bibitem{DAgostini:1993arp}
G.~D'Agostini, \textit{{On the use of the covariance matrix to fit correlated
  data}},  {\em Nucl. Instrum. Meth.} \textbf{A346} (1994) 306--311.

\bibitem{Amhis:2016xyh}
Y.~Amhis et~al., \textit{{Averages of $b$-hadron, $c$-hadron, and $\tau$-lepton
  properties as of summer 2016}},
  \href{https://arxiv.org/abs/1612.07233}{\texttt{arXiv:1612.07233}}.

\bibitem{Hamano:2008iba}
K.~Hamano, {\em {Measurement of Branching Fractions and Form Factor Parameters
  of $B \to Dl\nu$ and $B \to D^*l\nu$ Decays at BaBar}}.
\newblock PhD thesis, Victoria U., 2008.

\bibitem{1512299}
\textbf{Belle} Collaboration, \textit{{Precise determination of the CKM matrix
  element $\left| V_{cb}\right|$ with $\bar B^0 \to D^{*\,+} \, \ell^- \, \bar
  \nu_\ell$ decays with hadronic tagging at Belle}},  2017.
\newblock \url{https://www.hepdata.net/record/ins1512299}.

\bibitem{Grossman:1994ax}
Y.~Grossman and Z.~Ligeti, \textit{{The Inclusive $\bar B \to \tau \bar\nu X$
  decay in two Higgs doublet models}},  {\em Phys. Lett.} \textbf{B332} (1994)
  373--380,
  [\href{https://arxiv.org/abs/hep-ph/9403376}{\texttt{hep-ph/9403376}}].

\bibitem{Celis:2016azn}
A.~Celis, M.~Jung, X.-Q. Li, and A.~Pich, \textit{{Scalar contributions to
  $b\to c (u) \tau \nu$ transitions}},  {\em Phys. Lett.} \textbf{B771} (2017)
  168--179,
  [\href{https://arxiv.org/abs/1612.07757}{\texttt{arXiv:1612.07757}}].

\bibitem{Jezabek:1988iv}
M.~Jezabek and J.~H. Kuhn, \textit{{QCD Corrections to Semileptonic Decays of
  Heavy Quarks}},  {\em Nucl. Phys.} \textbf{B314} (1989) 1.

\bibitem{Czarnecki:1992zm}
A.~Czarnecki and S.~Davidson, \textit{{QCD corrections to the charged Higgs
  decay of a heavy quark}},  {\em Phys. Rev.} \textbf{D48} (1993) 4183--4187,
  [\href{https://arxiv.org/abs/hep-ph/9301237}{\texttt{hep-ph/9301237}}].

\bibitem{Czarnecki:1994bn}
A.~Czarnecki, M.~Jezabek, and J.~H. Kuhn, \textit{{Radiative corrections to
  $b\to c \tau \bar \nu_\tau$}},  {\em Phys. Lett.} \textbf{B346} (1995)
  335--341,
  [\href{https://arxiv.org/abs/hep-ph/9411282}{\texttt{hep-ph/9411282}}].

\bibitem{Grossman:1995yp}
Y.~Grossman, H.~E. Haber, and Y.~Nir, \textit{{QCD corrections to charged Higgs
  mediated $b\to c$ tau-neutrino decay}},  {\em Phys. Lett.} \textbf{B357}
  (1995) 630--636,
  [\href{https://arxiv.org/abs/hep-ph/9507213}{\texttt{hep-ph/9507213}}].

\bibitem{Li:2016vvp}
X.-Q. Li, Y.-D. Yang, and X.~Zhang, \textit{{Revisiting the one leptoquark
  solution to the $R(D^{(\star)})$ anomalies and its phenomenological
  implications}},  {\em JHEP} \textbf{08} (2016) 054,
  [\href{https://arxiv.org/abs/1605.09308}{\texttt{arXiv:1605.09308}}].

\bibitem{Alonso:2016oyd}
R.~Alonso, B.~Grinstein, and J.~Martin~Camalich, \textit{{The lifetime of the
  $B_c^-$ meson and the anomalies in $B\to D^{(*)}\tau\nu$}},  {\em Phys. Rev.
  Lett.} \textbf{118} (2017), no.~8 081802,
  [\href{https://arxiv.org/abs/1611.06676}{\texttt{arXiv:1611.06676}}].

\bibitem{Crivellin:2012ye}
A.~Crivellin, C.~Greub, and A.~Kokulu, \textit{{Explaining $B\to D\tau\nu$,
  $B\to D^*\tau\nu$ and $B\to \tau\nu$ in a 2HDM of type III}},  {\em Phys.
  Rev.} \textbf{D86} (2012) 054014,
  [\href{https://arxiv.org/abs/1206.2634}{\texttt{arXiv:1206.2634}}].

\bibitem{Celis:2012dk}
A.~Celis, M.~Jung, X.-Q. Li, and A.~Pich, \textit{{Sensitivity to charged
  scalars in $\boldsymbol{B\to D^{(*)}\tau\nu_\tau}$ and
  $\boldsymbol{B\to\tau\nu_\tau}$ decays}},  {\em JHEP} \textbf{01} (2013) 054,
  [\href{https://arxiv.org/abs/1210.8443}{\texttt{arXiv:1210.8443}}].

\bibitem{Tanaka:2012nw}
M.~Tanaka and R.~Watanabe, \textit{{New physics in the weak interaction of
  $\bar B\to D^{(*)}\tau\bar\nu$}},  {\em Phys. Rev.} \textbf{D87} (2013),
  no.~3 034028,
  [\href{https://arxiv.org/abs/1212.1878}{\texttt{arXiv:1212.1878}}].

\bibitem{Greljo:2015mma}
A.~Greljo, G.~Isidori, and D.~Marzocca, \textit{{On the breaking of Lepton
  Flavor Universality in B decays}},  {\em JHEP} \textbf{07} (2015) 142,
  [\href{https://arxiv.org/abs/1506.01705}{\texttt{arXiv:1506.01705}}].

\bibitem{Freytsis:2015qca}
M.~Freytsis, Z.~Ligeti, and J.~T. Ruderman, \textit{{Flavor models for $\bar{B}
  \to D^{(*)} \tau \bar{\nu}$}},  {\em Phys. Rev.} \textbf{D92} (2015), no.~5
  054018, [\href{https://arxiv.org/abs/1506.08896}{\texttt{arXiv:1506.08896}}].

\bibitem{Calibbi:2015kma}
L.~Calibbi, A.~Crivellin, and T.~Ota, \textit{{Effective Field Theory Approach
  to $b\to s\ell\ell^{(\prime)}$, $B\to K^{(*)}\nu\overline{\nu}$ and $B\to
  D^{(*)}\tau\nu$ with Third Generation Couplings}},  {\em Phys. Rev. Lett.}
  \textbf{115} (2015) 181801,
  [\href{https://arxiv.org/abs/1506.02661}{\texttt{arXiv:1506.02661}}].

\bibitem{Fajfer:2015ycq}
S.~Fajfer and N.~Košnik, \textit{{Vector leptoquark resolution of $R_K$ and
  $R_{D^{(*)}}$ puzzles}},  {\em Phys. Lett.} \textbf{B755} (2016) 270--274,
  [\href{https://arxiv.org/abs/1511.06024}{\texttt{arXiv:1511.06024}}].

\bibitem{Barbieri:2015yvd}
R.~Barbieri, G.~Isidori, A.~Pattori, and F.~Senia, \textit{{Anomalies in
  $B$-decays and $U(2)$ flavour symmetry}},  {\em Eur. Phys. J.} \textbf{C76}
  (2016), no.~2 67,
  [\href{https://arxiv.org/abs/1512.01560}{\texttt{arXiv:1512.01560}}].

\bibitem{Das:2016vkr}
D.~Das, C.~Hati, G.~Kumar, and N.~Mahajan, \textit{{Towards a unified
  explanation of $R_{D^{(\ast)}}$, $R_{K}$ and $(g-2)_{\mu}$ anomalies in a
  left-right model with leptoquarks}},  {\em Phys. Rev.} \textbf{D94} (2016)
  055034, [\href{https://arxiv.org/abs/1605.06313}{\texttt{arXiv:1605.06313}}].

\bibitem{Becirevic:2016yqi}
D.~Bečirević, S.~Fajfer, N.~Košnik, and O.~Sumensari, \textit{{Leptoquark
  model to explain the $B$-physics anomalies, $R_K$ and $R_D$}},  {\em Phys.
  Rev.} \textbf{D94} (2016), no.~11 115021,
  [\href{https://arxiv.org/abs/1608.08501}{\texttt{arXiv:1608.08501}}].

\bibitem{Gonzalez-Alonso:2017iyc}
M.~González-Alonso, J.~Martin~Camalich, and K.~Mimouni,
  \textit{{Renormalization-group evolution of new physics contributions to
  (semi)leptonic meson decays}},  {\em Phys. Lett.} \textbf{B772} (2017)
  777--785,
  [\href{https://arxiv.org/abs/1706.00410}{\texttt{arXiv:1706.00410}}].

\bibitem{Voloshin:1997zi}
M.~B. Voloshin, \textit{{Bound on V + A admixture in the $b\to c$ current from
  inclusive versus exclusive semileptonic decays of B mesons}},  {\em Mod.
  Phys. Lett.} \textbf{A12} (1997) 1823--1827,
  [\href{https://arxiv.org/abs/hep-ph/9704278}{\texttt{hep-ph/9704278}}].

\bibitem{Nierste:2008qe}
U.~Nierste, S.~Trine, and S.~Westhoff, \textit{{Charged-Higgs effects in a new
  $B \to D \tau \nu$ differential decay distribution}},  {\em Phys. Rev.}
  \textbf{D78} (2008) 015006,
  [\href{https://arxiv.org/abs/0801.4938}{\texttt{arXiv:0801.4938}}].

\bibitem{Sakaki:2013bfa}
Y.~Sakaki, M.~Tanaka, A.~Tayduganov, and R.~Watanabe, \textit{{Testing
  leptoquark models in $\bar B \to D^{(*)} \tau \bar\nu$}},  {\em Phys. Rev.}
  \textbf{D88} (2013), no.~9 094012,
  [\href{https://arxiv.org/abs/1309.0301}{\texttt{arXiv:1309.0301}}].

\bibitem{ForemanMackey:2012ig}
D.~Foreman-Mackey, D.~W. Hogg, D.~Lang, and J.~Goodman, \textit{{emcee: The
  MCMC Hammer}},  {\em Publ. Astron. Soc. Pac.} \textbf{125} (2013) 306--312,
  [\href{https://arxiv.org/abs/1202.3665}{\texttt{arXiv:1202.3665}}].

\end{thebibliography}\endgroup

\end{document}